\documentclass[aip,
 amsmath,amssymb,
]{revtex4-2}

\usepackage{graphicx}
\usepackage{dcolumn}
\usepackage{bm}

\usepackage[utf8]{inputenc}
\usepackage[T1]{fontenc}
\usepackage{mathptmx}
\usepackage{etoolbox}
\usepackage{xcolor}
\usepackage{subfig}
\usepackage{hyperref}

\makeatletter
\def\@email#1#2{%
 \endgroup
 \patchcmd{\titleblock@produce}
  {\frontmatter@RRAPformat}
  {\frontmatter@RRAPformat{\produce@RRAP{*#1\href{mailto:#2}{#2}}}\frontmatter@RRAPformat}
  {}{}
}%
\makeatother
\begin{document}

\title{Turbulent transport regimes in the tokamak boundary and operational limits}

\author{M. Giacomin}
\email{maurizio.giacomin@epfl.ch}
\author{P. Ricci}%
 \affiliation{Ecole Polytechnique F\'{e}d\'{e}rale de Lausanne (EPFL), Swiss Plasma Center (SPC), CH-1015 Lausanne, Switzerland}
 
\begin{abstract}
Two-fluid, three-dimensional, flux-driven, global, electromagnetic turbulence simulations carried out by using the GBS (Global Braginskii Solver) code are used to identify the main parameters controlling turbulent transport in the tokamak boundary and to delineate an electromagnetic phase space of edge turbulence. Four turbulent transport regimes are identified: (i) a regime of fully developed turbulence appearing at intermediate values of collisionality and $\beta$, with turbulence driven by resistive ballooning modes, related to the L-mode operation of tokamaks, (ii) a regime of reduced turbulent transport at low collisionality and large heat source, with turbulence driven by drift-waves, related to a high-density H-mode regime, (iii) a regime of extremely large turbulent transport at high collisionality, which is associated with the crossing of the density limit, and (iv) a regime above the ideal ballooning limit at high $\beta$, with global modes affecting the dynamics of the entire confined region, which can be associated with the crossing of the $\beta$ limit. 
The transition from the reduced to the developed turbulent transport regime is associated here with the H-mode density limit and an analytical scaling law for maximum edge density achievable in H-mode is obtained.
Analogously, analytical scaling laws for the crossing of the L-mode density and $\beta$ limits are provided and compared to the results of GBS simulations.  

\end{abstract}

\maketitle

\section{Introduction}

Identifying the main parameters controlling plasma turbulence in the tokamak boundary and understanding the physical mechanisms behind the transition between the various turbulent regimes is of major importance for the design and operation of future magnetic fusion devices.
In fact, the limits that restraint the operational space of tokamaks, such as the density limit,~\cite{greenwald1988,greenwald2002} as well as important phenomena that play a fundamental role in determining the overall performance of a tokamak, such as the L-H transition,~\cite{Wagner1982} strongly depend on the nonlinear turbulent plasma dynamics in the tokamak boundary.

Several regimes of tokamak operation with different confinement properties have been achieved experimentally in the past years.~\cite{viezzer2018} Among these regimes, the high confinement mode (H-mode)~\cite{Wagner1982}  has been chosen as ITER baseline scenario.
The H-mode is achieved above a certain power threshold and is characterized by an edge transport barrier that is responsible for
steep edge temperature and density gradients compared to the low
confinement mode (L-mode).
The maximum density achievable in H-mode is denoted as the H-mode density limit. A back transition from the H-mode to the L-mode is observed when the density exceeds the H-mode density limit.
The H-mode density limit differs from the standard H-L transition caused by a reduction of the power crossing the separatrix below the H-mode power threshold, since the H-mode density limit can be reached even at values of the power crossing the separatrix that are larger than the H-mode power threshold.~\cite{mertens2000,borrass2004,bernert2014h}

The tokamak plasma density cannot exceed a certain threshold also in L-mode operation.
A widely-used empirical scaling of the maximum line-averaged density that can be achieved was obtained by Greenwald in 1988,~\cite{greenwald1988}
\begin{equation}
\label{eqn:greenwald}
    n_{GW}[10^{20} \text{m}^{-3}] = \frac{I_p [\text{MA}]}{\pi a[\text{m}]^2}\,,
\end{equation}
where $n_{GW}$, known as Greenwald density, is the predicted maximum line-averaged density, $I_p$ the plasma current and $a$ the plasma minor radius.
The Greenwald density limit, also denoted as the L-mode density limit, is a hard limit, namely its crossing leads to the onset of magnetohydrodynamics (MHD) modes, performance degradation and a plasma disruption.~\cite{greenwald1988,greenwald2002}
Despite the fact that both the L-mode and the H-mode density limits are experimentally observed to occur at similar density values, the H-mode density limit differs from the L-mode density limit. In fact, the H-mode density limit is usually a soft limit since plasma operation can be continued in L-mode after the H-L transition.~\cite{huber2013,bernert2014h}

In addition to the density limit, various MHD instabilities restrain  the operational space of tokamaks. Among these, the ideal ballooning instability, which occurs at large pressure gradient values, imposes the maximum value of $\beta$ that can be achieved in tokamaks.~\cite{wesson1978,wesson1985}
The $\beta$ limit is a hard limit.
Indeed, large-scale modes develop over the entire plasma when the $\beta$ limit is exceeded, leading to a plasma disruption.


A theoretical description of the different turbulent transport regimes at the tokamak edge and their link to the tokamak operational limits was first provided in Refs.~\onlinecite{Scott1997,rogers1997,rogers1998}, based on flux-tube two-fluid turbulent simulations.
In particular, a phase space of edge turbulence, including the L-H transition, the ideal MHD limit and the Greenwald density limit, was derived in Ref.~\onlinecite{rogers1998} in terms of the MHD parameter
\begin{equation}
    \label{eqn:alpha_mhd_first}
    \alpha_\text{MHD} = - R_0 q^2 \frac{\mathrm{d}\beta}{\mathrm{d}r} \simeq R_0 q^2 \frac{\beta}{L_p}
\end{equation}
and of the diamagnetic parameter 
\begin{equation}
    \alpha_d = \sqrt{\frac{m_i c_s \tau_e}{0.51 m_e 4 \pi^2 q^2 R_0}}\Bigl(\frac{R_0}{L_p}\Bigr)^{1/4}\,,
\end{equation}
where $R_0$ is the tokamak major radius, $q$ is the safety factor, $r$ denotes the cross-field direction, $c_s$ is the sound speed, $\tau_e$ is the electron collisional time and $L_p$ is the edge pressure gradient length. 
In the phase space described in Ref.~\onlinecite{rogers1998}, the L-H transition occurs at high values of $\alpha_{\text{MHD}}$ and $\alpha_d$, the ideal MHD limit is reached at large values of $\alpha_{\text{MHD}}$, independently of the $\alpha_d$ value, and the density limit is crossed at low $\alpha_d$, i.e. high collisionality, and finite $\alpha_\text{MHD}$.  
The crossing of the density limit described in Refs.~\onlinecite{rogers1997,rogers1998} is associated with a regime of catastrophically large turbulent transport in the tokamak edge resulting from nonlinear electromagnetic effects.
Therefore,  Ref.~\onlinecite{rogers1998} claims that no density limit can be retrieved in the electrostatic case, underlining the key role played by electromagnetic fluctuations. 
Similarly, Ref.~\onlinecite{eich2021} has linked the crossing of the density limit to a transition from an electrostatic to an electromagnetic ballooning regime, again underlining the important role played by electromagnetic fluctuations in the density limit, even though a different mechanism than the one proposed in Refs.~\onlinecite{rogers1997,rogers1998}, which is based on a transition between the driving linear modes, is invoked.
In contrast, the theoretical works reported in Refs.~\onlinecite{hajjar2018,singh2021} argue that the key parameter controlling turbulent transport at the tokamak edge is the collisionality, rather than $\beta$, and suggests that a regime of large turbulent transport, compatible with the crossing of the density limit, can be achieved even at low $\beta$.
Also the electromagnetic gyrokinetic tokamak boundary simulations described in Refs.~\onlinecite{mandell2020,mandell2021electromagnetic} show a weak effect of electromagnetic perturbations on turbulence and equilibrium profiles, thus suggesting a secondary role played by $\beta$ on the edge turbulent transport. 

A recent theoretical investigation based on flux-driven, two-fluid, three-dimensional electrostatic turbulent simulations, carried out with the GBS (Global Braginskii Solver) code and using the Boussinesq approximation, has identified three different turbulent transport regimes in the tokamak edge.~\cite{giacomin2020transp}
These include a regime of reduced turbulent transport at low collisionality and large heat source, with turbulence driven by the Kelvin-Helmholtz instability, a regime of developed turbulent transport at intermediate values of collisionality and heat source, with turbulence driven by resistive ballooning modes, and a regime of extremely large turbulent transport at high collisionality and low heat source, with turbulence still driven by resistive ballooning modes, associated with the crossing of the density limit.
Despite being in the electrostatic limit, and therefore neglecting any effect due to electromagnetic fluctuations, the simulations reported in Ref.~\onlinecite{giacomin2020transp} show the presence of a density limit crossing.
In a recent work, the result of Ref.~\onlinecite{giacomin2020transp} has been leveraged to derive a theory-based scaling law for the density limit that shows a better agreement with a multi-machine database than the Greenwald empirical scaling.~\cite{giacomin2022density}

In this work, we extend the results presented in Ref.~\onlinecite{giacomin2020transp} by leveraging a set of three-dimensional, flux-driven, two-fluid electromagnetic turbulence simulations, carried out with the GBS code.\cite{giacomin2021gbs}
With respect to the simulations in Ref.~\onlinecite{giacomin2020transp}, we consider here simulations that include electromagnetic effects and avoid the use of the Boussinesq approximation.
We derive an electromagnetic phase space of edge turbulence where four turbulent transport regimes are identified: (i) a regime of fully developed turbulence appearing at intermediate values of collisionality and $\beta$, with turbulence driven by resistive ballooning modes, which we associate with the L-mode operation of tokamaks, (ii) a regime of reduced turbulent transport and improved confinement at low collisionality and large heat source, with turbulence driven by the drift-wave instability, associated with the H-mode regime in high-density conditions, (iii)  a regime of extremely large turbulent transport at high collisionality, low heat source and realistic values of $\beta$, which is associated with the crossing of the density limit, and (iv)  a regime above the ideal ballooning limit at high $\beta$, with global modes developing on the entire confined region that leads to a total loss of plasma and heat, which can be associated with the crossing of the $\beta$ limit. 
We find that the density limit crossing is independent of $\beta$ (for values of $\beta$ below the $\beta$ limit), thus pointing out the secondary role played by electromagnetic fluctuations on turbulent transport while approaching the density limit. This finding is in contrast to Refs.~\onlinecite{Scott1997},~\onlinecite{rogers1998}~and~\onlinecite{eich2021}, while it confirms the result of Ref.~\onlinecite{giacomin2020transp}.
In addition, the transition from the drift-wave regime to the resistive ballooning regime is associated with the H-mode density limit, and an analytical scaling of the maximum density that can be achieved in the H-mode operating conditions before causing the H-L back transition is derived. 

The present paper is organized as follows. The physical model considered in this work is summarized in Sec.~\ref{sec:phys_model}, while an overview of the simulation results is presented in Sec.~\ref{sec:overview}, where different turbulent transport regimes are identified from GBS simulations. In Sec.~\ref{sec:phase_space}, an electromagnetic phase space of edge turbulence is derived and analytical estimates of the edge pressure gradient length are provided.
The transitions among the different regimes identified here are then analyzed in Sec.~\ref{sec:transitions}, where analytical estimates of the H-mode density limit, L-mode density limit and $\beta$ limit are provided and compared to the results of GBS simulations. 
A comparison of the edge phase space derived in this work with past investigations is presented in Sec.~\ref{sec:comparison}.
The conclusions follow in Sec.~\ref{sec:conclusions}.

\section{Physical model}\label{sec:phys_model}

The physical model considered here is based on the drift-reduced Braginskii model~\cite{Zeiler1997} implemented in GBS.~\cite{giacomin2021gbs}
For simplicity, the coupling to the neutral dynamics is neglected, although implemented in GBS.~\cite{mancini2021}
The validity of a drift-reduced fluid model is limited to the regime of electron mean free path shorter than the parallel connection length, $\lambda_e \ll L_\parallel \simeq 2\pi q R$, and perpendicular scale lengths of the dominant modes larger than the ion Larmor radius, $k_\perp \rho_i \ll 1$.
These conditions are usually verified in the tokamak boundary of L-mode discharges. On the other hand, the steep pedestal temperature in H-mode discharges leads, most often, to  collisionality values such that $\lambda_e \gtrsim L_\parallel$ and turbulence driven by unstable modes with $k_\perp \rho_i\sim 1$,~\cite{jenko2001,dickinson2012} whose exhaustive characterization  requires to account for kinetic effects. 
On the other hand, H-mode discharges at high density feature collisionality values sufficiently large that fluid models can be applied for their description.
For example, a H-mode TCV discharge near the H-mode density limit~\cite{pau2020} with edge electron density $n_e\simeq 5\times 10^{19}$~m$^{-3}$ and edge electron temperature $T_e\simeq 150$~eV yields $\lambda_e/L_\parallel \simeq 0.1$, which justifies the use of a fluid model in the proximity of the H-mode density limit.
In addition, the physical model neglects the bootstrap current, thus excluding the peeling instability from the system. While the bootstrap current plays an important role in pedestal stability and edge-localized modes (see, e. g., Refs.~\onlinecite{wilson1999,snyder2004}), its effect is expected to be negligible in the high density and high collisionality regimes considered in this work.
The use of drift-reduced fluid model restricts therefore our study to L-mode discharges and H-mode discharges at high density and high collisionality.

The model equations considered in the present work are
\begin{align}
\label{eqn:density}
\frac{\partial n}{\partial t} =& -\frac{\rho_*^{-1}}{B}\bigl[\phi,n\bigr]+\frac{2}{B}\Bigl[C(p_e)-nC(\phi)\Bigr] 
-\nabla_{\parallel}(n v_{\parallel e}) + D_n\nabla_{\perp}^2 n +s_n\, ,\\
\label{eqn:vorticity}
\frac{\partial \Omega}{\partial t} =& -\frac{\rho_*^{-1}}{B}\nabla \cdot [\phi,\boldsymbol{\omega}] - \nabla \cdot \bigl( v_{\parallel i}\nabla_\parallel \boldsymbol{\omega}\bigr) + B^2\nabla_{\parallel}j_{\parallel} + 2B C(p_e + \tau p_i) \nonumber\\
&+ \frac{B}{3}C(G_i) + D_{\Omega}\nabla_\perp^2 \Omega\, ,\\
\label{eqn:electron_velocity}
\frac{\partial U_{\parallel e}}{\partial t} =& -\frac{\rho_*^{-1}}{B}[\phi,v_{\parallel e}]  + \frac{m_i}{m_e}\Bigl(\nu j_\parallel+\nabla_\parallel\phi-\frac{1}{n}\nabla_\parallel p_e-0.71\nabla_\parallel T_e -\frac{2}{3n}\nabla_\parallel G_e\Bigr)\nonumber\\ 
&- v_{\parallel e}\nabla_\parallel v_{\parallel e}+ D_{v_{\parallel e}}\nabla_\perp^2 v_{\parallel e}\,, \\
\label{eqn:ion_velocity}
\frac{\partial v_{\parallel i}}{\partial t} =& -\frac{\rho_*^{-1}}{B}\bigl[\phi,v_{\parallel i}\bigr] - v_{\parallel i}\nabla_\parallel v_{\parallel i} - \frac{1}{n}\nabla_\parallel(p_e+\tau p_i)
+ \frac{4}{3n}\eta_{0,i}\nabla^2_\parallel v_{\parallel i} + D_{v_{\parallel i}}\nabla_\perp^2 v_{\parallel i}\, ,\\
\label{eqn:electron_temperature}
\frac{\partial T_e}{\partial t} =& -\frac{\rho_*^{-1}}{B}\bigl[\phi,T_e\bigr] - v_{\parallel e}\nabla_\parallel T_e 
+ \frac{2}{3}T_e\Bigl[0.71\nabla_\parallel v_{\parallel i}-1.71\nabla_\parallel v_{\parallel e}
+0.71 (v_{\parallel i}-v_{\parallel e})\frac{\nabla_\parallel n}{n}\Bigr] \nonumber \\
&+ \frac{4}{3}\frac{T_e}{B}\Bigl[\frac{7}{2}C(T_e)+\frac{T_e}{n}C(n)-C(\phi)\Bigr] 
+ \chi_{\parallel e}\nabla_\parallel^2 T_e + D_{T_e}\nabla_\perp^2 T_e + s_{T_e}\,,\\
\label{eqn:ion_temperature}
\frac{\partial T_i}{\partial t} =& -\frac{\rho_*^{-1}}{B}\bigl[\phi,T_i\bigr] - v_{\parallel i}\nabla_\parallel T_i 
+ \frac{4}{3}\frac{T_i}{B}\Bigl[C(T_e)+\frac{T_e}{n}C(n)-C(\phi)\Bigr] - \frac{10}{3}\tau\frac{T_i}{B}C(T_i) \nonumber \\ 
&+ \frac{2}{3}T_i(v_{\parallel i}-v_{\parallel e})\frac{\nabla_\parallel n}{n} -\frac{2}{3}T_i\nabla_\parallel v_{\parallel e} 
+ \chi_{\parallel i}\nabla_\parallel^2 T_i + D_{T_i}\nabla_\perp^2 T_i + s_{T_i}\,,
\end{align}
which are coupled to Poisson and Ampère equations, 
\begin{align}
\label{eqn:poisson}
\nabla \cdot \bigl( n \nabla_\perp \phi\bigr) &= \ \Omega-\tau\nabla_\perp^2 p_i\,,\\
\label{eqn:ampere}
\biggl( \nabla_\perp^2 - \frac{\beta_{e0}}{2}\frac{m_i}{m_e}n\biggr)v_{\parallel e}&= \ \nabla_\perp^2 U_{\parallel e} - \frac{\beta_{e0}}{2}\frac{m_i}{m_e}n v_{\parallel i} + \frac{\beta_{e0}}{2}\frac{m_i}{m_e} \overline{j}_\parallel\,,
\end{align}
where $\Omega = \nabla\cdot\boldsymbol{\omega} = \nabla \cdot (n \nabla_\perp\phi + \tau \nabla_\perp p_i)$ is the scalar vorticity and $U_{\parallel e} = v_{\parallel e} + m_i \psi/m_e$ is the sum of electron inertia and electromagnetic induction contributions.
We highlight that, in contrast to the physical model considered in Ref.~\onlinecite{giacomin2020transp}, here we include electromagnetic effects by solving Ampère equation (see Eq.~\eqref{eqn:ampere}) and we avoid the use of the Boussinesq approximation in the vorticity and Poisson equations (see Eqs.~\eqref{eqn:vorticity}~and~\eqref{eqn:poisson}). 

In Eqs.~\eqref{eqn:density}-\eqref{eqn:ampere} and in the following, GBS normalized units are used. 
In particular, $n$, $T_e$ and $T_i$ are normalized to the reference values $n_0$, $T_{e0}$ and $T_{i0}$, respectively.
The electron and ion parallel velocities, $v_{\parallel e}$ and $v_{\parallel i}$, are normalized to the reference sound speed $c_{s0}=\sqrt{T_{e0}/m_i}$. The magnetic field is normalized to its modulus at the tokamak axis, $B_T$. The electrostatic potential, $\phi$, is normalized to $T_{e0}/e$, and $\psi$ is normalized to $\rho_{s0} B_T$, with $\rho_{s0}= c_{s0}/\Omega_{ci}$ the reference ion sound Larmor radius. Perpendicular lengths are normalized to $\rho_{s0}$ and parallel lengths are normalized to the tokamak major radius $R_0$. Time is normalized to $R_0/c_{s0}$.
The dimensionless parameters appearing in the model equations are the normalized ion sound Larmor radius, $\rho_* = \rho_{s0}/R_0$, the ion to electron temperature ratio, $\tau = T_{i0}/T_{e0}$, the normalized electron and ion parallel thermal conductivities, 
\begin{equation}
\label{eqn:parallel_cond}
    \chi_{\parallel e} = \chi_{\parallel e 0} T_e^{5/2} = \biggl(\frac{1.58}{\sqrt{2\pi}}\frac{m_i}{\sqrt{m_e}}\frac{(4\pi\epsilon_0)^2}{e^4}\frac{c_{s0}}{R_0}\frac{T_{e0}^{3/2}}{\lambda n_0}\biggr)T_e^{5/2}
\end{equation}
and
\begin{equation}
    \chi_{\parallel i} = \chi_{\parallel i 0} T_i^{5/2} = \biggl(\frac{1.94}{\sqrt{2\pi}}\sqrt{m_i}\frac{(4\pi\epsilon_0)^2}{e^4}\frac{c_{s0}}{R_0}\frac{T_{e0}^{3/2}\tau^{5/2}}{\lambda n_0}\biggr)T_i^{5/2}\,,
\end{equation}
the reference electron plasma $\beta$, 
\begin{equation}
\beta_{e0}=2\mu_0 \frac{n_0 T_{e0}}{B_T^2}\,,    
\end{equation}
and the normalized Spitzer resistivity, $\nu = e^2n_0R_0/(m_ic_{s0}\sigma_\parallel) = \nu_0 T_e^{-3/2}$, with 
\begin{align}
\sigma_\parallel =& \biggl(1.96\frac{n_0 e^2 \tau_e}{m_e}\biggr)n=\biggl(\frac{5.88}{4\sqrt{2\pi}}\frac{(4\pi\epsilon_0)^2}{e^2}\frac{ T_{e0}^{3/2}}{\lambda\sqrt{m_e}}\biggr)T_e^{3/2},\\
\label{eqn:resistivity}
\nu_0 =&\frac{4\sqrt{2\pi}}{5.88}\frac{e^4}{(4\pi\epsilon_0)^2}\frac{\sqrt{m_e}R_0n_0\lambda}{m_i c_{s0} T_{e0}^{3/2}},
\end{align}
where $\lambda$ is the Coulomb logarithm.
The gyroviscous terms are given by
\begin{align}
    G_i&=-\eta_{0i}\Bigl[2\nabla_\parallel v_{\parallel i}+\frac{1}{B}C(\phi)+\frac{\tau}{n B}C(p_i)\Bigr]\,,\\
    G_e&=-\eta_{0e}\Bigl[2\nabla_\parallel v_{\parallel e} +\frac{1}{B}C(\phi)-\frac{1}{nB}C(p_e)\Bigr]\,,
\end{align}
where $\eta_{0i}=0.96 T_{i0}\tau_i/(m_i R_0 c_{s0})$ and $\eta_{0e}=0.96 T_{e0}\tau_e/(m_e R_0 c_{s0})$. 
These dimensionless parameters depend on the values of the reference quantities that are usually evaluated at the separatrix. 

The spatial operators appearing in Eqs.~\eqref{eqn:density}--\eqref{eqn:poisson} are the $\mathbf{E}\times\mathbf{B}$ convective term $\bigl[\phi,f\bigr]=\mathbf{b}\ \cdot\ (\nabla \phi \times \nabla f)$, the curvature operator $C(f)=B\bigl[\nabla \times (\mathbf{b}/B)\bigr]/2\cdot \nabla f$, the perpendicular Laplacian operator ${\nabla_\perp^2 f=\nabla\cdot\bigl[(\mathbf{b}\times\nabla f)\times\mathbf{b}\bigr]}$ and the parallel gradient operator $\nabla_\parallel f=\mathbf{b}\cdot\nabla f + \bigl[\psi, f\bigr]/B$, where $\mathbf{b}=\mathbf{B}/B$ is the unit vector of the (unperturbed) magnetic field and $\bigl[\psi, f\bigr]/B$ is the electromagnetic flutter contribution. 
The toroidally symmetric equilibrium magnetic field is written in terms of the poloidal magnetic flux $\Psi$, normalized to $\rho_{s0}^2B_T$, as
\begin{equation}
\label{eqn:magnetic_field}
    \mathbf{B}=\pm\nabla\varphi+\rho_*\nabla\varphi\times\nabla\Psi ,
\end{equation}
where $\varphi$ is the toroidal angle. The plus (minus) sign in Eq.~\eqref{eqn:magnetic_field} refers to the direction of the toroidal magnetic field with the ion-$\nabla B$ drift pointing upwards (downwards).
The differential operators are discretized on a non-field-aligned $(R,\phi,Z)$ cylindrical grid by means of a fourth-order finite difference scheme, where $R$ is the radial distance from the tokamak symmetry axis and $Z$ is the vertical direction.

The source terms in the density and temperature equations, $s_n$ and $s_T$, are added to fuel and heat the plasma, and they are analytical functions of $\Psi(R,Z)$, independent of the toroidal angle:
\begin{align}
    \label{eqn:density_source}
    s_n &= s_{n0} \exp\biggl(-\frac{\bigl(\Psi(R,Z)-\Psi_{n}\bigr)^2}{\Delta_n^2}\biggr),\\   
    \label{eqn:temperature_source}
    s_T &= \frac{s_{T0}}{2}\biggl[\tanh\biggl(-\frac{\Psi(R,Z)-\Psi_{T}}{\Delta_T}\biggr)+1\biggr], 
\end{align}
where $\Psi_n$ and $\Psi_T$ are flux surfaces located inside the last closed flux surface (LCFS). The density source is localized around the flux surface $\Psi_n$, close to the separatrix, and mimics the ionization process, while the temperature source extends throughout the entire core region and mimics the ohmic heating. 
We define the total density and temperature source integrated over the area inside the LCFS as
\begin{equation}
    S_n=\int_{A_{\text{LCFS}}} \rho_* s_n(R,Z)\,\mathrm{d}R\mathrm{d}Z
\end{equation}
and
\begin{equation}
    S_T=\int_{A_{\text{LCFS}}} \rho_* s_T(R,Z)\,\mathrm{d}R\mathrm{d}Z\,,
\end{equation}
where the factor $\rho_*$ appears from our normalization choices. 
Analogously, we define the electron pressure source, proportional to the electron power source, as $S_p=\int_{A_{\text{LCFS}}} \rho_* s_p\,\mathrm{d}R\mathrm{d}Z$, with $s_p=n s_{T_e} + T_e s_n$ and $s_{T_e}$ the electron temperature source.
More details on the physical models and on its numerical implementation in GBS, as well as on the boundary conditions, are reported in Ref.~\onlinecite{giacomin2021gbs}.

\section{Overview of the simulation results}\label{sec:overview}

We now describe the results of the GBS electromagnetic simulations considered here, which have been carried out with the following dimensionless parameters: $\rho_*^{-1}=500$, $a/R_0\simeq 0.3$, $s_{n0}=0.3$, $\Delta_n = 800$, $\Delta_T = 720$, $\chi_{\parallel e0} = 10$, $\chi_{\parallel i0} = 1$,  upward ion-$\nabla B$ drift direction, $s_{T0}= \{ 0.15, 0.3, 0.6\}$, $\nu_0=\{0.05, 0.1, 0.2, 0.6, 10\}$, and various values of $\beta_{e0}$ ranging from $10^{-6}$ to $5\times 10^{-3}$. 
The magnetic equilibrium is the same as in Ref.~\onlinecite{giacomin2020transp}, namely it is analytically obtained in the infinite aspect-ratio limit by solving the Biot-Savart law for a current density with a Gaussian distribution centered at the tokamak axis, mimicking the plasma current, and an additional current filament outside the simulation domain to produce the X-point. The value of the plasma current and the width of its Gaussian distribution are chosen to have a safety factor $q_0 \simeq 1$ at the tokamak axis and $q_{95} \simeq 4$ at the tokamak edge.

In order to reduce the computational cost of the present simulations, the value of $\chi_{\parallel e 0}$ has been reduced by approximately an order of magnitude with respect to typical values in the tokamak boundary. Consequently, the parallel heat flux due to the plasma convection is significantly larger than the parallel heat flux due to conduction, i.e. $n T_e v_{\parallel e} \gg \chi_{\parallel e} \nabla_\parallel T_e$. 
On the other hand, the parallel heat conduction is usually larger than the parallel heat convection in experiments. In fact, by considering typical values of electron density and electron temperature at the separatrix of a TCV discharge ($n_0\simeq 10^{19}$~m$^{-3}$ and $T_{e0}\simeq 30$~eV), the parallel heat flux due to conduction is approximately $\chi_{\parallel e} \nabla_\parallel T_e \simeq \chi_{\parallel e} T_e/(q R_0) \sim 10$~MW/m$^2$ and is larger than the parallel heat transport due to convection, $n T_e v_{\parallel e} \simeq n T_e c_{s}\sim 1$~MW/m$^2$.
In the present paper, while the main analysis and comparison to simulation results is done in the convection limit, the theoretical scaling laws we derive are provided in both convection and conduction limits, thus allowing for a future comparison with experimental data.

The analysis described in the following is carried out when the simulations are in a global turbulent quasi-steady state resulting from the balance among the sources in the closed flux surface region, turbulence that transports plasma and heat from the core to the scrape-off layer (SOL), and the losses at the vessel.
The equilibrium component of a quantity $f$, denoted as $\bar{f}$, is evaluated by taking the time and toroidal average of $f$, while the fluctuating component is defined as $\tilde{f}=f-\bar{f}$.
The flux-aligned coordinate system $(\nabla\Psi,\nabla\chi,\nabla\varphi)$ is used in the analysis, where $\nabla \psi$ denotes the direction perpendicular to flux surfaces, $\nabla \varphi$ denotes the toroidal direction and $\nabla\chi = \nabla\varphi\times\nabla\Psi$.

\begin{figure}
    \centering
    \subfloat[]{\includegraphics[height=0.15\textheight]{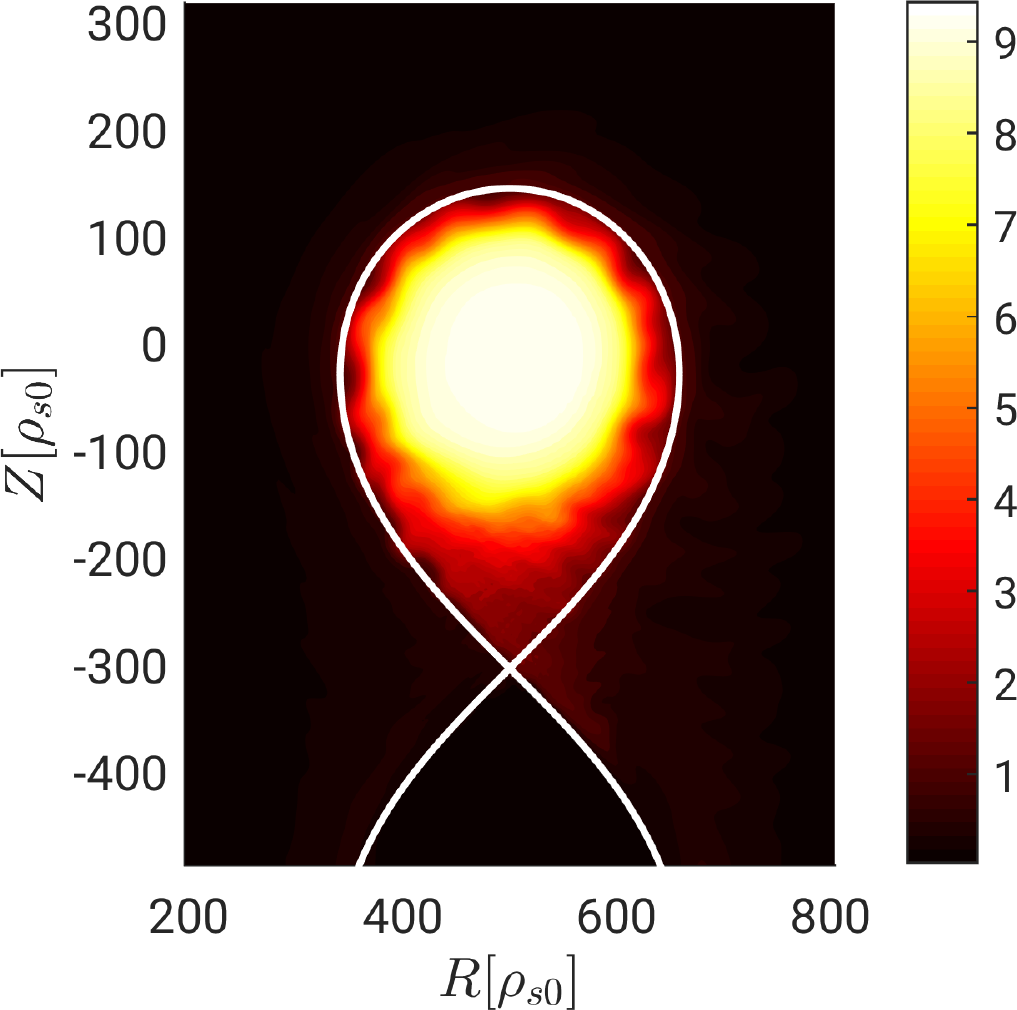}}\,
    \subfloat[]{\includegraphics[height=0.15\textheight]{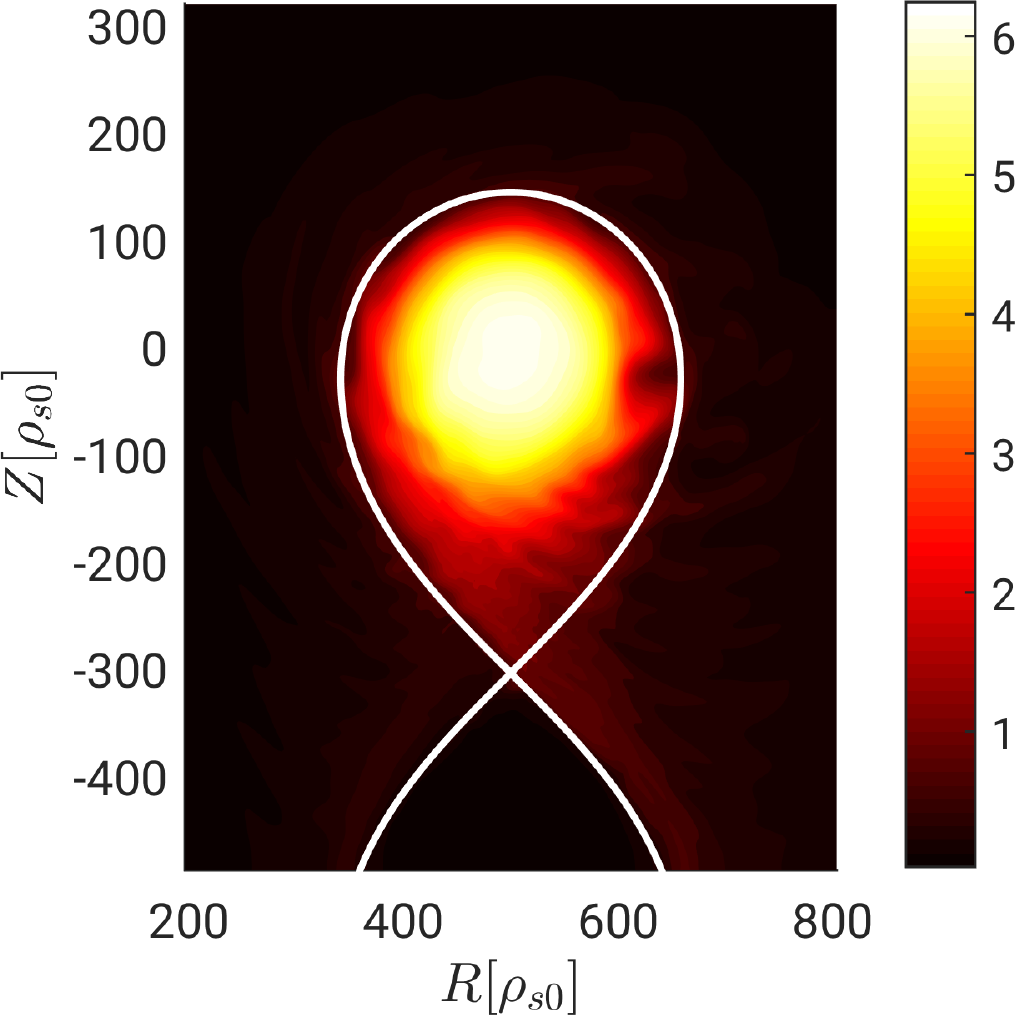}}\,
    \subfloat[]{\includegraphics[height=0.15\textheight]{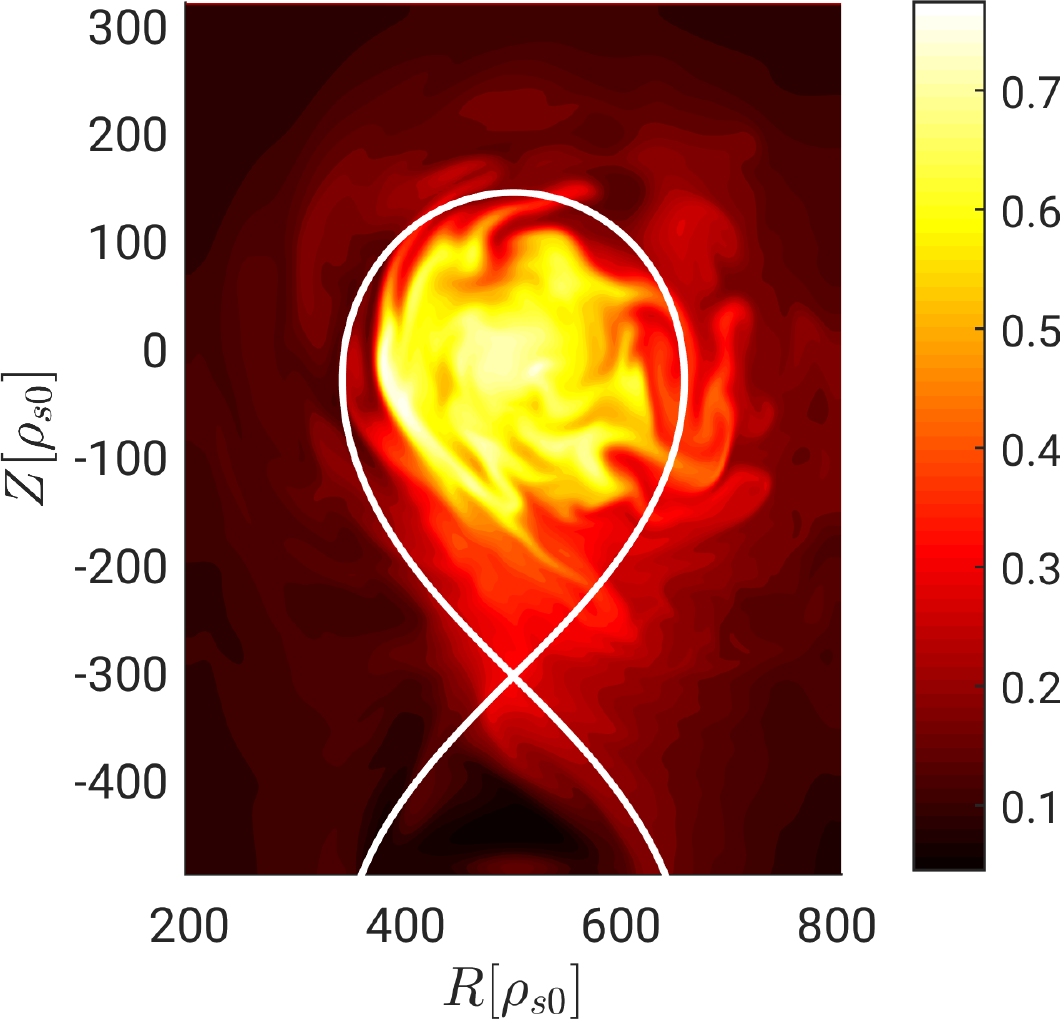}}\,
    \subfloat[]{\includegraphics[height=0.15\textheight]{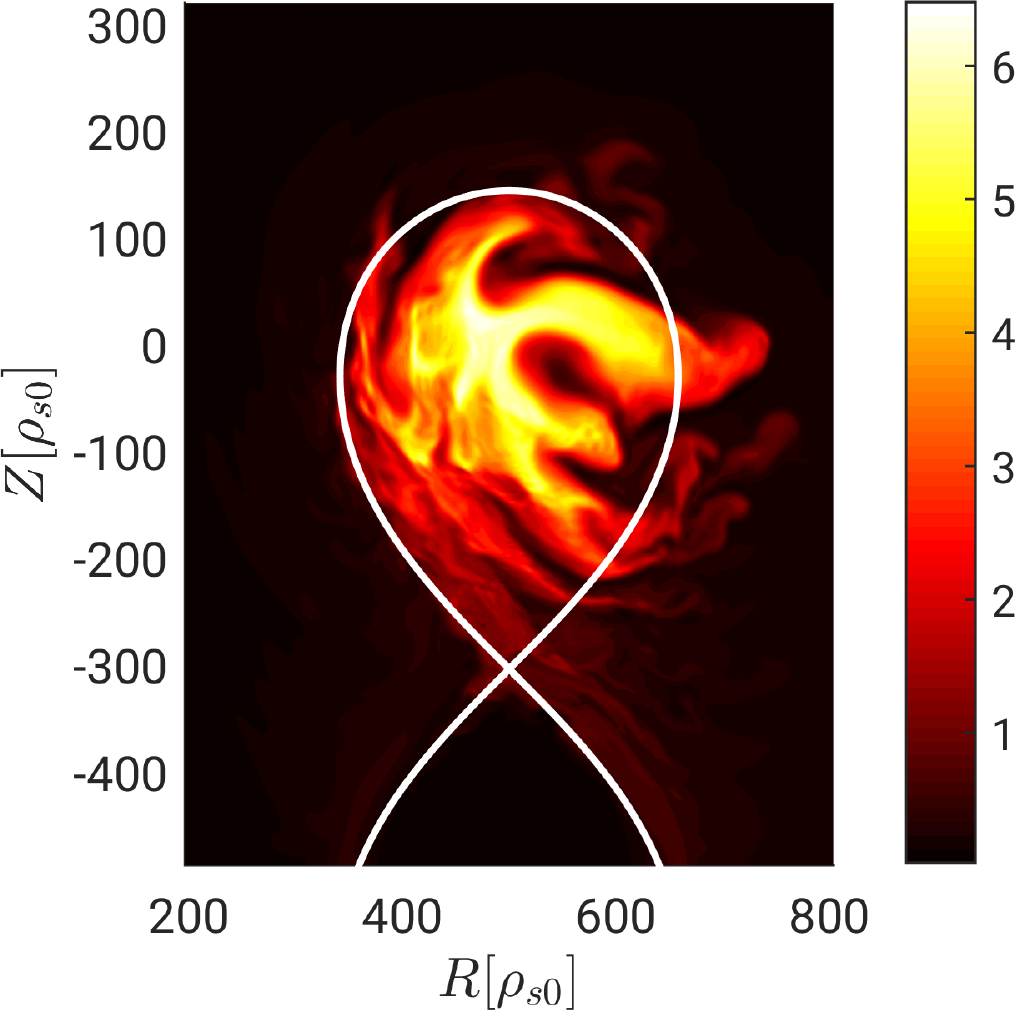}}
    \caption{Typical snapshots of density in the suppressed transport regime, $\nu_0=0.05$ and $\beta_{e0}=10^{-6}$ (a), in the developed transport regime, $\nu_0=0.2$ and $\beta_{e0}=10^{-4}$ (b), above the density limit, $\nu_0=10$ and $\beta_{e0}=10^{-4}$ (c), and above the $\beta$ limit, $\nu_0=0.2$ and $\beta_{e0}= 4 \times 10^{-4}$ (d). The same value of $s_{T0} = 0.3$ is considered in these simulations.  The white line represents the separatrix.  }
    \label{fig:density_2d}
\end{figure}

In Fig.~\ref{fig:density_2d}, typical snapshots of the plasma density for the electromagnetic simulations that avoid the Boussinesq approximation are shown at various values of $\nu_0$ and $\beta_{e0}$, corresponding to the different turbulent transport regimes observed in our simulations.
In contrast to Ref.~\onlinecite{giacomin2020transp}, where three electrostatic turbulent transport regimes are described, four electromagnetic regimes can be identified here. 

\begin{figure}
    \centering
    \subfloat[]{\includegraphics[width=0.3\textwidth]{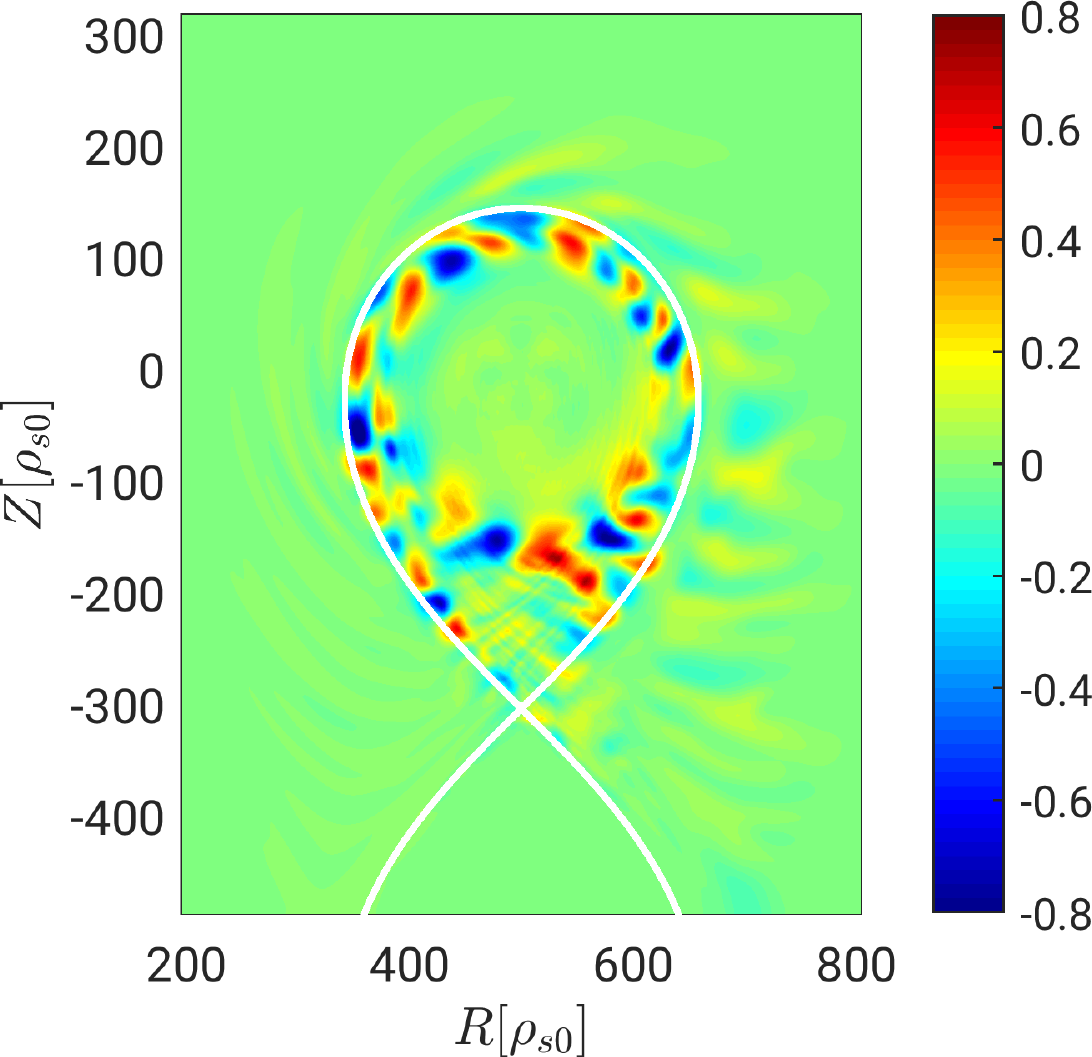}}\quad
    \subfloat[]{\includegraphics[width=0.3\textwidth]{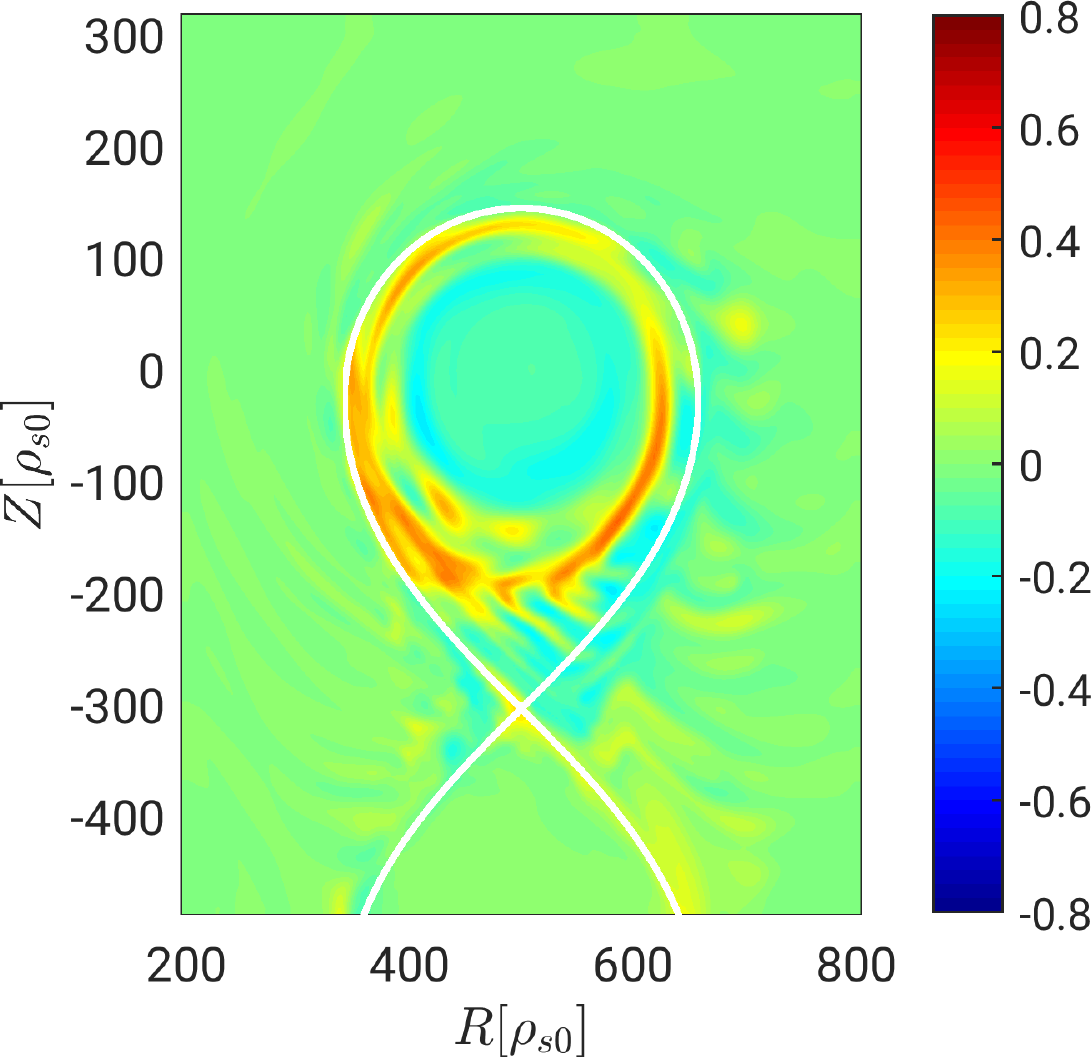}}\quad
    \subfloat[]{\includegraphics[width=0.3\textwidth]{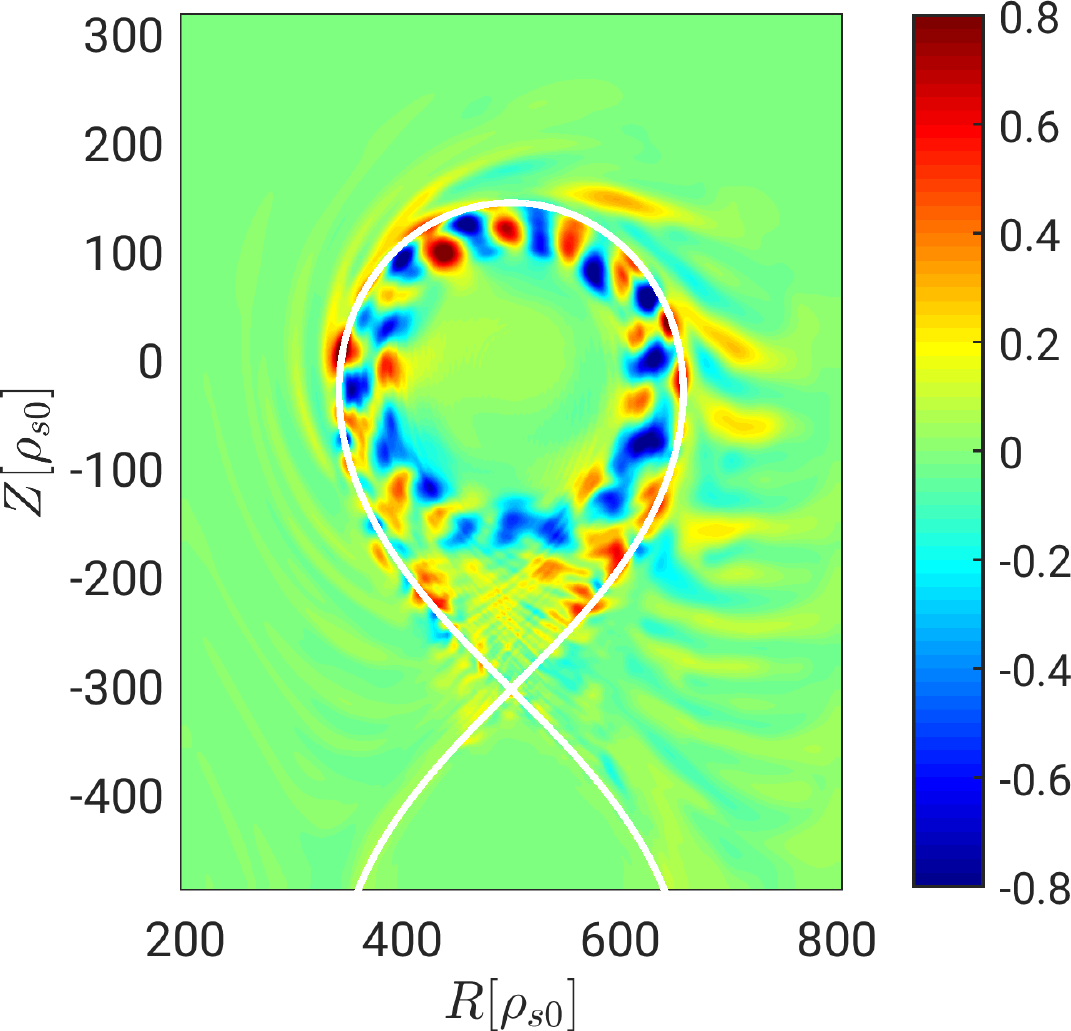}}
    \caption{Typical density fluctuations of the simulation with $\nu_0=0.05$, $s_{T0}=0.3$ and $\beta_{e0}=10^{-6}$ (a). The panels (b) and (c) show a typical snapshot of density fluctuations when the drift-wave instability (the term $\nabla_\parallel p_e/n + 0.71 \nabla T_e$ in Eq.~\eqref{eqn:electron_velocity}) or the drive of the Kelvin-Helmholtz instability (the term $\nabla \cdot [\phi,\boldsymbol{\omega}]$ in Eq.~\eqref{eqn:vorticity}) is removed from the dynamics. }
    \label{fig:DW_KH}
\end{figure}

At very low values of collisionality and high heat source, a reduced turbulence regime, characterized by a steep edge pressure profile, is observed.
Turbulence in this regime is mainly driven by the drift-wave instability.
This is revealed by performing a test similar to the one carried out in Ref.~\onlinecite{giacomin2020transp}, whose results are shown in Fig.~\ref{fig:DW_KH}.
Namely, for the simulation with $\nu_0=0.05$, $s_{T0}=0.3$ and $\beta_{e0}=10^{-6}$,  drift-waves are removed from the system by zeroing out the term $\nabla_\parallel p_e/n + 0.71 \nabla_\parallel T_e$ in Eq.~\eqref{eqn:electron_velocity}. 
Fig.~\ref{fig:DW_KH} shows that density fluctuations vanish when the drift-waves are removed from the dynamics, clearly indicating that, in the low collisionality and high heat source regime, turbulence is mainly driven by the drift-wave instability.
On the other hand, only a weak effect on density fluctuations is observed when the drive of Kelvin-Helmholtz instability (the term $\nabla \cdot [\phi,\boldsymbol{\omega}]$ in Eq.~\eqref{eqn:vorticity}) is removed from the system, thus excluding Kelvin-Helmholtz from being the primary instability in these simulations (Fig.~\ref{fig:DW_KH}~(c)).
This contrasts with the findings in Ref.~\onlinecite{giacomin2020transp}, where the reduced transport regime found at low collisionality and large values of heat source is characterized by turbulence driven by the Kelvin-Helmholtz instability, showing considerably larger values of the $\mathbf{E}\times \mathbf{B}$ shear than the typical values observed in the electromagnetic simulations.

We note that the differences between the present simulations and the ones in Ref.~\onlinecite{giacomin2020transp} persist also at low $\beta$.
In fact, these differences are due to the use of the Boussinesq approximation in Ref.~\onlinecite{giacomin2020transp}, $\nabla \cdot (n\nabla_\perp \phi + \tau \nabla_\perp p_i) \simeq n (\nabla_\perp^2\phi + \tau \nabla_\perp^2 T_i/n)$, which is avoided here.
This shows that, although the Boussinesq approximation is commonly used to simulate tokamak boundary turbulence,~\cite{bodi2011,Ricci2012,giacomin2020snow} its validity becomes questionable in the region across the separatrix,~\cite{stegmeir2019} where steep density gradients can form, especially in the regime of reduced turbulent transport, where the use of the Boussinesq approximation significantly affects the character of the driving instability.

We remark that the theoretical work proposed in Ref.~\onlinecite{rogers1998} associates the transition with the H-mode to a transition to a regime where edge turbulence is mostly driven by the drift-wave instability.
Here, we link this regime to a high density H-mode and we associate the transition from the drift-wave regime with the resistive ballooning regime to a H-mode density limit, as described in Sec.~\ref{sec:transitions}. 
We also note that a regime dominated by drift-wave turbulence has been recently found also in gyro-fluid simulations and associated with the I-mode regime observed in tokamaks.~\cite{manz2020}

A test similar to the one in Fig.~\ref{fig:DW_KH} shows that the resistive ballooning instability dominates over the drift-wave instability at intermediate values of collisionality and $\beta$. 
In the resistive ballooning regime, the $\mathbf{E}\times \mathbf{B}$ shear plays only a minor role and no transport barrier forms across the separatrix. Similarly to Ref.~\onlinecite{giacomin2020transp}, this regime can be associated with the standard L-mode of tokamak operation.
In contrast to the drift-wave regime, the use of the Boussinesq approximation in the resistive ballooning regime has a weak effect on turbulence and equilibrium profiles.

\begin{figure}
    \centering
    \subfloat[]{\includegraphics[height=0.17\textheight]{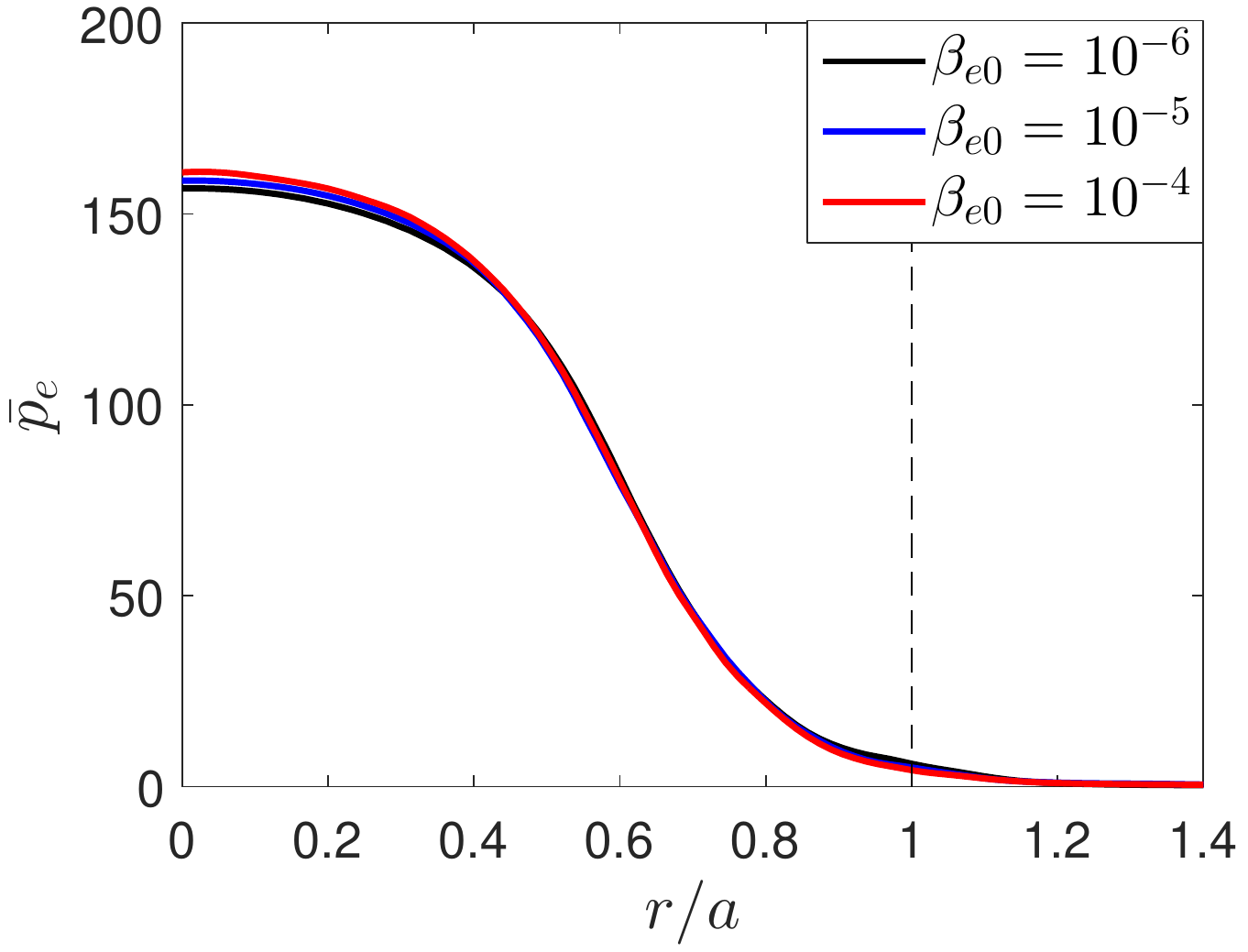}}\,
    \subfloat[]{\includegraphics[height=0.165\textheight]{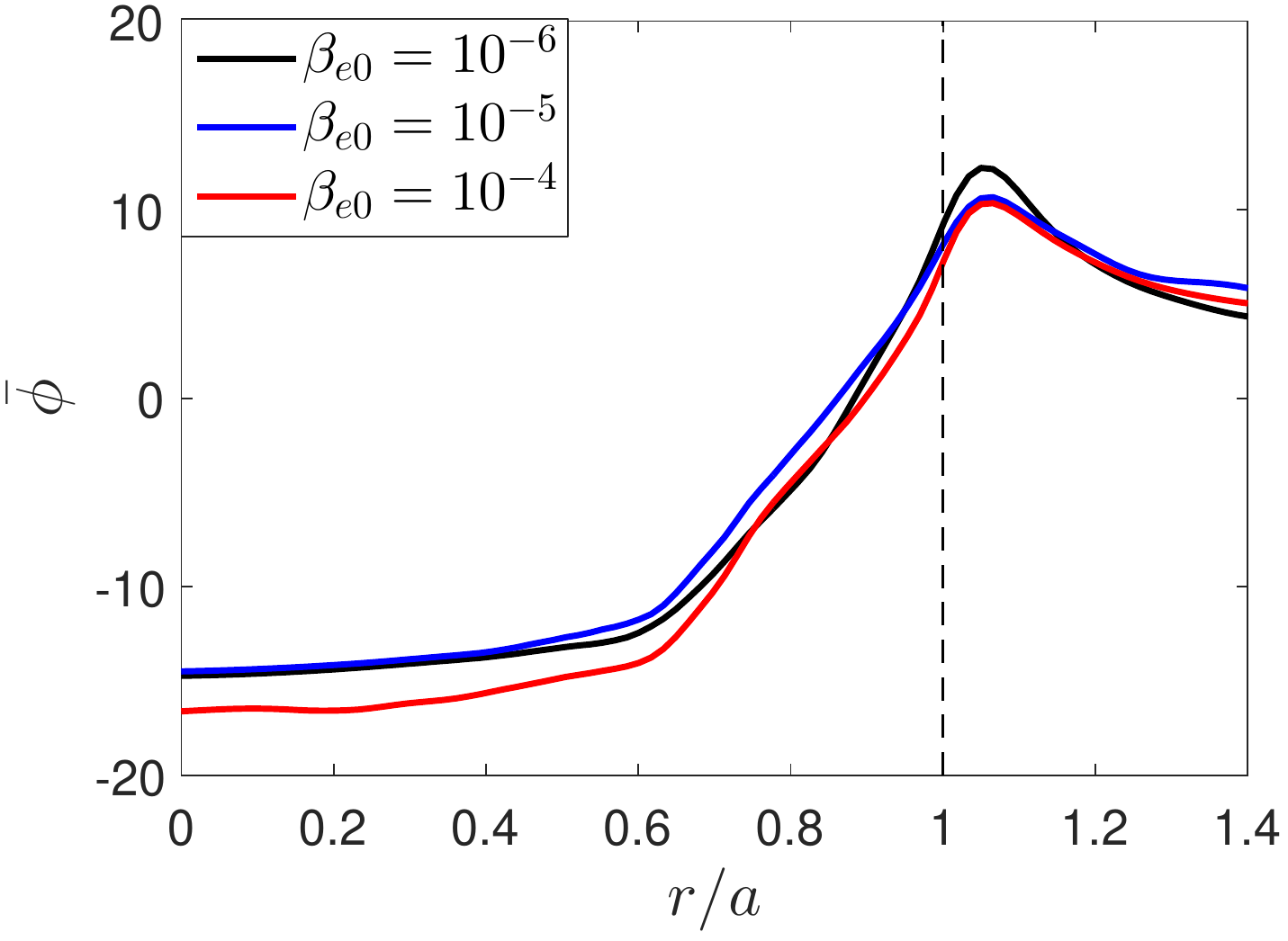}}\,
    \subfloat[]{\includegraphics[height=0.165\textheight]{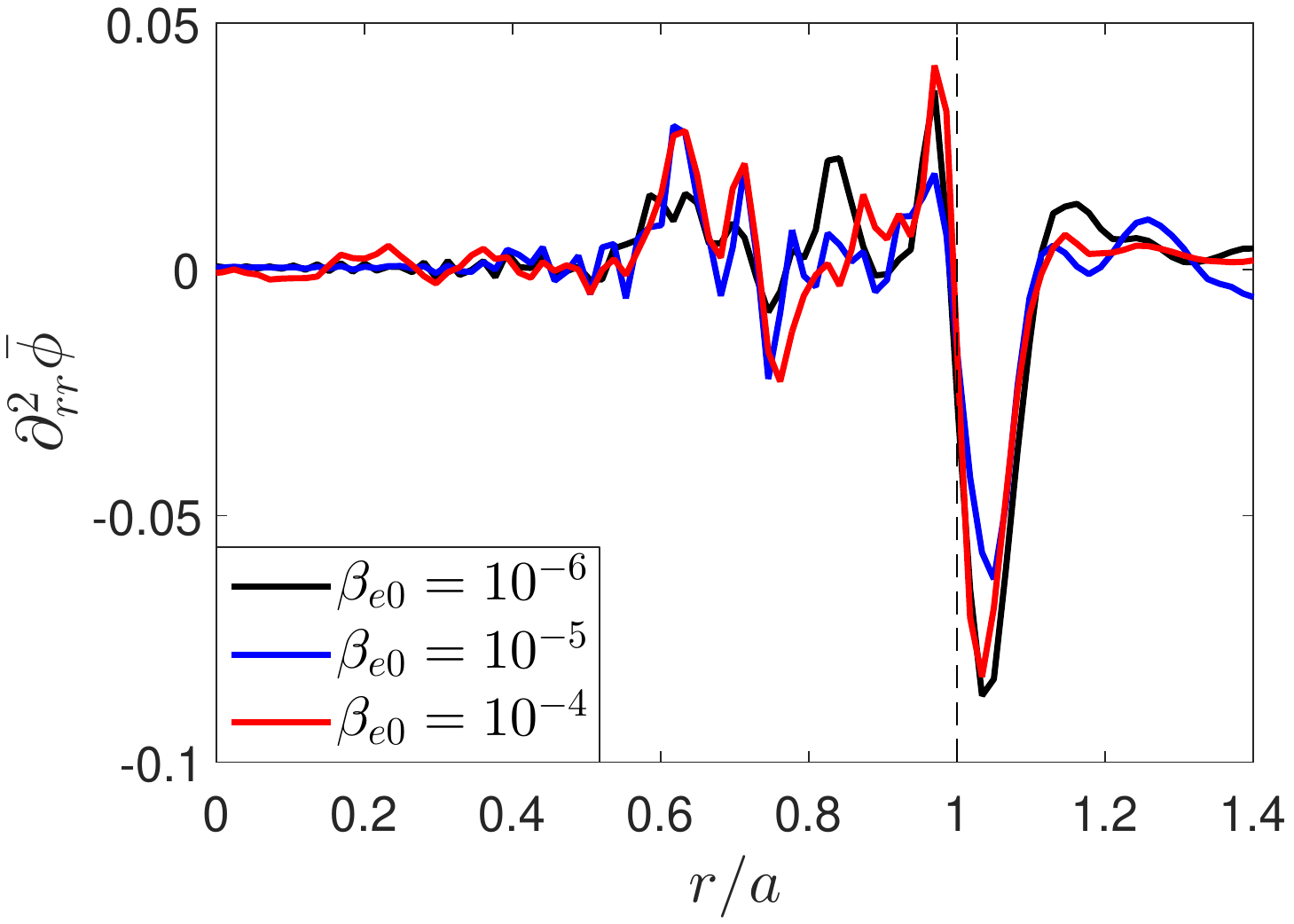}}
    \caption{Equilibrium radial profiles at the outboard midplane of electron pressure (a), electrostatic potential (b) and $\mathbf{E}\times\mathbf{B}$ shear (c) for simulations in the resistive ballooning regime at various values of $\beta_{e0}$ with $s_{T0}=0.3$ and $\nu_0=0.2$. The vertical dashed line represents the position of the separatrix. }
    \label{fig:radial_profiles}
\end{figure}

The effect of electromagnetic fluctuations on the resistive ballooning regime is investigated in Fig.~\ref{fig:radial_profiles}, where the equilibrium radial profiles of electron pressure, electrostatic potential and $\mathbf{E}\times\mathbf{B}$ shear at the outboard midplane are shown for the simulations at $\nu_0=0.2$, $s_{T0}=0.3$ and three different values of $\beta_{e0}$, below the $\beta$ limit, covering a range of two orders of magnitude.
The radial profiles show a very weak dependence on $\beta_{e0}$, suggesting that turbulent transport is weakly affected by this parameter at realistic values of $\beta_{e0}$.
In addition, turbulent transport due to the electromagnetic flutter is found to be negligible in all the simulations considered in the present work. 
We conclude that electromagnetic effects play only a minor role on edge turbulent transport in the resistive ballooning regime.
This result is in agreement with recent gyrokinetic simulations of the tokamak boundary, which show a weak dependence of equilibrium profiles on $\beta$.~\cite{mandell2020,mandell2021electromagnetic}

At large values of $\nu_0$, turbulent eddies extend throughout the entire core region (see Fig.~\ref{fig:density_2d}~(c)) and turbulent transport is extremely large.
Consequently, the equilibrium pressure and temperature gradients near the separatrix collapse.  
This regime of very large turbulent transport and flat pressure and temperature profiles, which is retrieved at high density, is linked to a regime beyond the density limit, in agreement with the result of electrostatic simulations presented in Ref.~\onlinecite{giacomin2020transp}. 
At these large values of collisionality, the Boussinesq approximation and electromagnetic perturbations have no effect on turbulence and equilibrium profiles.

Finally, at large values of $\beta_{e0}$, the ideal branch of the ballooning instability overcomes the resistive one.~\cite{mosetto2013} Consequently, ideal ballooning modes become the main instability driving turbulence.
The onset of the ideal ballooning instability generates global modes that affect the entire confined region, as shown in Fig.~\ref{fig:density_2d}~(d), eventually leading to a loss of confinement that corresponds to a plasma disruption.
This regime, characterized by global modes and large values of $\beta$, is associated with a regime beyond the $\beta$ limit. 

In the theoretical study proposed in Refs.~\onlinecite{rogers1997,rogers1998}, the crossing of the density limit is described as the result of the presence of electromagnetic fluctuations that inhibit the formation of sheared flows, which provide a saturation mechanism for resistive ballooning modes.  In our simulations, however, the density limit is observed also at very low values of $\beta_{e0}$ and even in the electrostatic limit. 
In fact, at high values of edge collisionality, simulations show negligible sheared flows near the separatrix at any value of $\beta_{e0}$, while nonlinear saturation of the pressure fluctuation amplitude is provided by the gradient removal mechanism,~\cite{Ricci2013,Ricci2008} rather than by a nonlinear mechanism associated with the presence sheared flows.
On the other hand, the presence of a density limit at low values of $\beta_{e0}$ observed in the simulations presented here is in agreement with the theoretical investigations of Ref.~\onlinecite{hajjar2018}, arguing that the edge collisionality is the main key parameter that controls turbulent transport and density limit crossing, independently of the $\beta$ value.
We note that an increase of turbulent transport with $\beta_{e0}$ is reported in Ref.~\onlinecite{halpern2013} only for values of $\beta$ that are above the $\beta$ limit.

\section{Electromagnetic phase space of boundary turbulence}\label{sec:phase_space}

\begin{figure}
    \centering
    \includegraphics[scale=0.6]{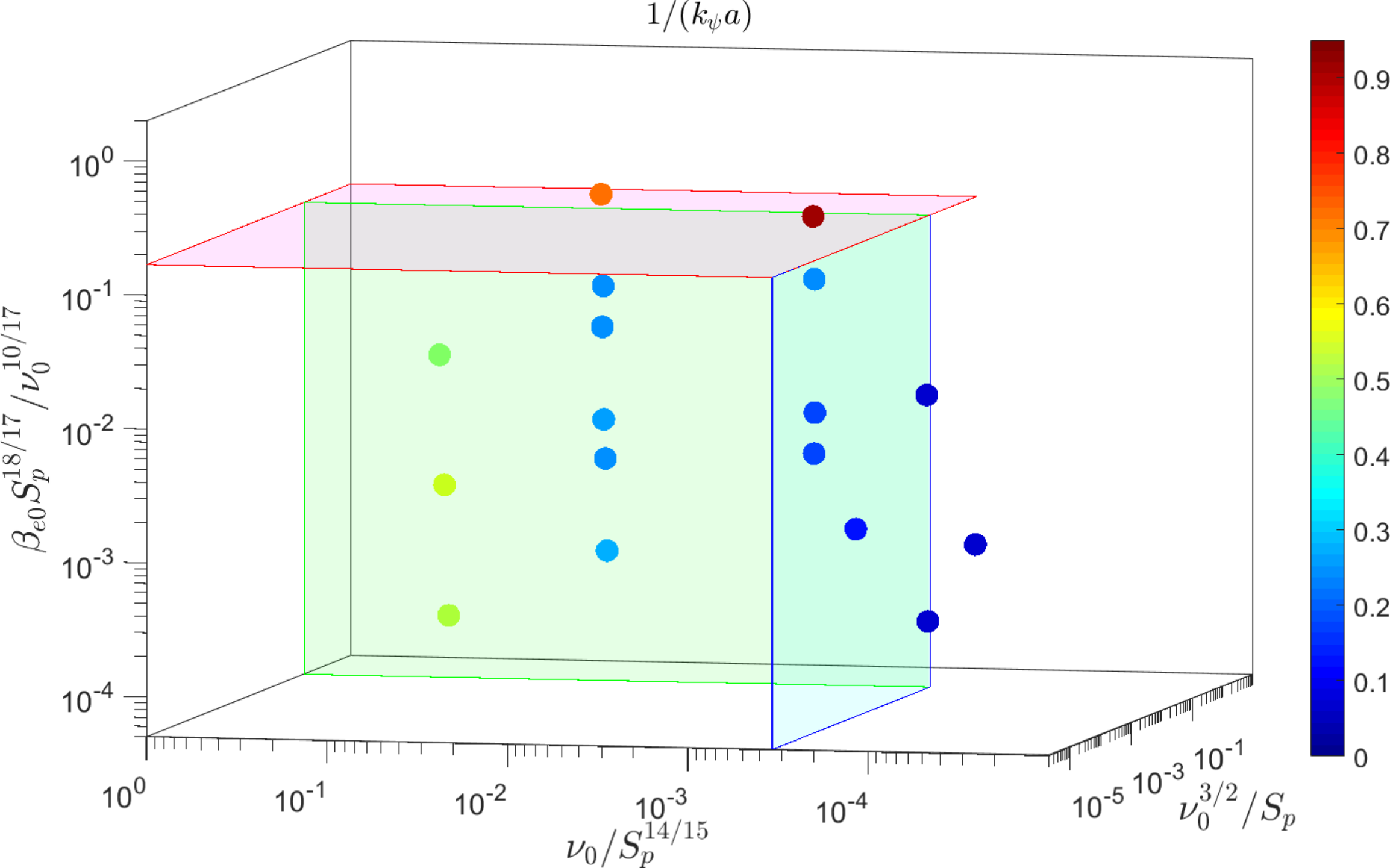}
    \caption{Time and toroidal average of radial extension of the largest turbulent eddies normalized to the tokamak minor radius, $1/(k_\psi a)$, for all the simulations considered in the present work, as a function of the parameters $\beta_{e0} S_p^{18/17}/\nu_0^{10/17}$, $\nu_0/S_p^{15/14}$ and $\nu_0^{3/2}/S_p$, which define our three-dimensional edge turbulence phase space. 
    The light blue plane corresponds to the H-mode density limit (see Eq.~\eqref{eqn:hmode_dl_bou}), the green plane to the L-mode density limit (see Eq.~\eqref{eqn:res_lim_fin}) and red plane to the $\beta$ limit (see Eq.~\eqref{eqn:beta_limit}), respectively. The density and $\beta$ boundaries delimit the parameter space where the plasma is confined. }
    \label{fig:phase_space}
\end{figure}

\begin{figure}
    \centering
    {\subfloat[]{\includegraphics[height=0.25\textheight]{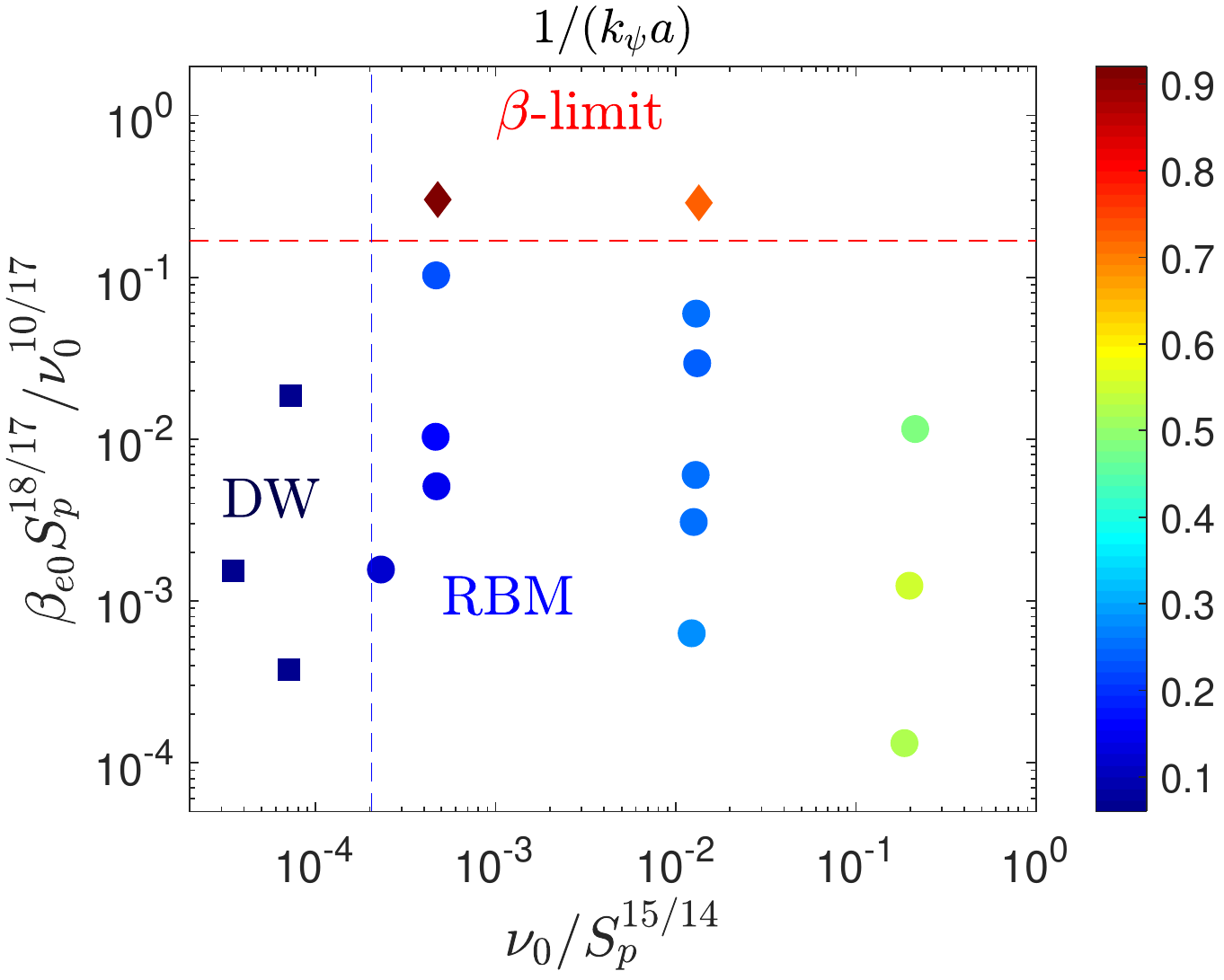}}}\quad
    {\subfloat[]{\includegraphics[height=0.25\textheight]{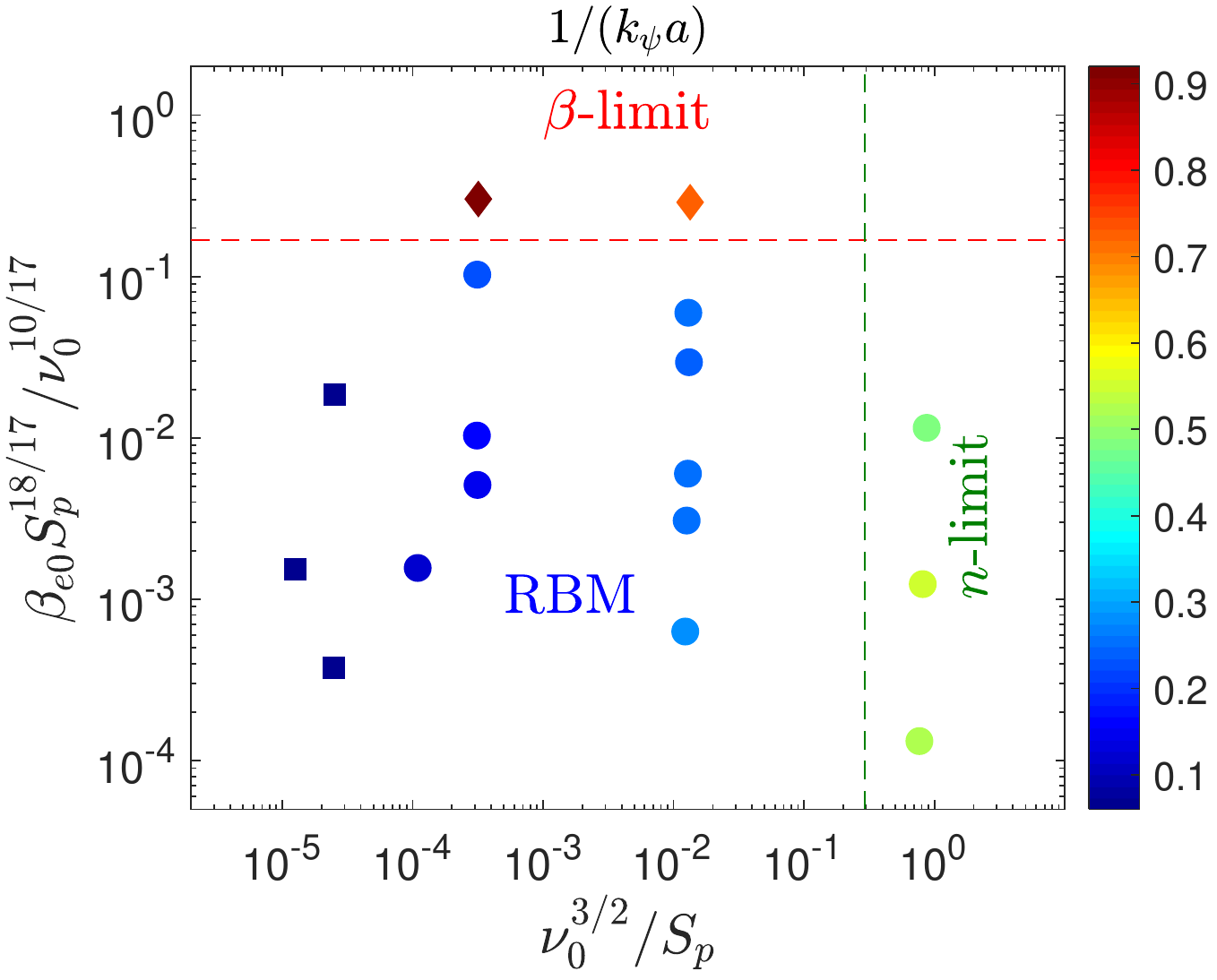}}}
    \caption{Projection of the three dimensional phase space in Fig.~\ref{fig:phase_space} onto the plane defined by the parameters $\nu_0/S_p^{15/14}$ and $\beta_{e0} S_p^{18/17}/\nu_0^{10/17}$ (a) and by  $\nu_0^{3/2}/S_p$ and $\beta_{e0} S_p^{18/17}/\nu_0^{10/17}$ (b). The dashed blue line represents the H-mode density limit (see Eq.~\eqref{eqn:hmode_dl_bou}), the dashed green line the L-mode density limit (see Eq.~\eqref{eqn:res_lim_fin}) and the red dashed line the $\beta$ limit (see Eq.~\eqref{eqn:beta_limit}). }
    \label{fig:projection_phase_space}
\end{figure}

The electromagnetic phase space of boundary turbulence derived from GBS simulations is outlined in Fig.~\ref{fig:phase_space}. The time and toroidal average of the radial extension of the largest turbulent eddies, expressed as $1/(k_\psi a)$ with $k_\psi$ the radial wave vector, is shown for all the simulations considered in the present work and is indicated by the colorbar. 
Four main regions are identified in the parameter space of Fig.~\ref{fig:phase_space}: (i) a region where the radial extension of turbulent eddies is significantly smaller than the tokamak minor radius, $1/(k_\psi a) \ll 1$, and turbulence is mainly driven by the drift-wave instability; (ii) a region where $1/(k_\psi a) \simeq 0.1$ and turbulence is mainly driven by resistive ballooning modes; (iii) a region at high $\nu_0$ characterized by very large turbulent transport, poor plasma confinement and $1/(k_\psi a) \simeq 0.5$, associated with the crossing of a density limit; and (iv) a region at large values of $\beta_{e0}$ characterized by large scale modes affecting the whole core plasma, $1/(k_\psi a)\simeq 1$, and associated with a regime beyond the $\beta$ limit.
Projections of the three-dimensional phase space in Fig.~\ref{fig:phase_space} onto two dimensional planes are shown in Fig.~\ref{fig:projection_phase_space}.

The three parameters controlling turbulent transport in Fig.~\ref{fig:phase_space} are $\nu_0/S_p^{14/15}$, $\nu_0^{3/2}/S_p$ and $\beta_{e0} S_p^{18/17}/\nu_0^{10/17}$, which are associated with the H-mode density limit transition, to the L-mode density limit crossing and to the transition between the resistive ballooning and the ideal ballooning regimes, respectively. These limits are derived in Sec.~\ref{sec:transitions}.
We note that the controlling parameters are written in terms of the dimensionless parameters $\nu_0$, $S_p$ and $\beta_{e0}$, which are the ones varied across the simulation scan presented in Sec.~\ref{sec:overview}.

We also remark that the regime of tokamak operation is bounded by the density and $\beta$ limits, and therefore it includes simulations with turbulence being driven either by resistive ballooning modes or drift-waves.
In this section, we focus on the two instabilities that appear when plasma is confined and we provide an analytical estimate of the equilibrium pressure gradient length near the separatrix.

\subsection{Drift-wave turbulence}

An analytical estimate of $L_p$ when drift-waves constitute the turbulence drive can be derived by following a procedure similar to the one described in Ref.~\onlinecite{giacomin2020transp} for the resistive ballooning regime, which is based on a balance between the cross-field turbulent heat flux at the LCFS, $q_{\psi} \simeq \overline{\tilde{p}_e\partial_\chi\tilde{\phi}}$, obtained from a quasi-linear non-local theory, and the heat source integrated over the poloidal plane inside the LCFS, i.e. 
\begin{equation}
    \label{eqn:flux_balance}
    S_p \simeq \oint_{\text{LCFS}} q_{\psi}\,\mathrm{d}l\,.
\end{equation}

The quantity $\partial_\chi\tilde{\phi}$ is estimated from the linearized electron pressure equation,
\begin{equation}
    \label{eqn:lin_pressure}
    \partial_t \tilde{p}_e \sim -\rho_*^{-1} \partial_\psi \bar{p}_e \partial_\chi\tilde{\phi}\,,
\end{equation}
which is obtained by summing and linearizing Eqs.~\eqref{eqn:density}~and~\eqref{eqn:electron_temperature}, where only the leading order terms are considered. 
The time derivative in Eq.~\eqref{eqn:lin_pressure} is now approximated by the growth rate of the driving drift-wave instability, where $L_{p,\text{\scriptsize{DW}}}$ is the equilibrium pressure gradient length across the LCFS in the drift-wave regime, while the radial derivative of $\bar{p}_e$ is approximated as $\partial_\psi \bar{p}_e \simeq \bar{p}_e/L_{p,\text{\scriptsize{DW}}}$. 
This leads to
\begin{equation}
    \label{eqn:transport_intermediate_first}
    q_{\psi,\text{\scriptsize{DW}}}\sim \rho_*\gamma_\text{\scriptsize{DW}}\frac{\tilde{p}_e^2}{\bar{p}_e}L_{p,\text{\scriptsize{DW}}}\,,
\end{equation}
where $\bar{p}_e$ is the equilibrium pressure evaluated at the LCFS.
The fluctuating electron pressure is obtained by assuming that the growth of the linearly unstable modes saturates when the instability drive is removed from the system, i.e. $k_{\psi,\text{\scriptsize{DW}}} \tilde{p}_e \sim \bar{p}_e/L_{p,\text{\scriptsize{DW}}}$,~\cite{Ricci2013,Ricci2008} with $k_{\psi,\text{\scriptsize{DW}}}\simeq\sqrt{k_{\chi,\text{\scriptsize{DW}}}/L_{p,\text{\scriptsize{DW}}}}$, as derived from the non-local analysis outlined in Ref.~\onlinecite{Ricci2008}.
Therefore, Eq.~\eqref{eqn:transport_intermediate_first} can be written as
\begin{equation}
    \label{eqn:transport_intermediate}
    q_{\psi,\text{\scriptsize{DW}}}\sim \rho_*\frac{\gamma_\text{\scriptsize{DW}}}{k_{\chi,\text{\scriptsize{DW}}}}\bar{n}\bar{T}_e\,.
\end{equation}

We remark that the effects of sheared flows are neglected in Eq.~\eqref{eqn:transport_intermediate} and in the following, although sheared flows are included in GBS simulations. 
This approximation is motivated by the result of the analysis reported in the Appendix, which shows a negligible effect of sheared flows on the drift-wave instability. 
The analysis of the drift-wave instability carried out in Ref.~\onlinecite{ricci2010} within the limit of negligible sheared flows leads to $\gamma_\text{\scriptsize{DW}}\simeq 0.12 \bar{T}_e^{1/2}/(\rho_* L_{p,\text{\scriptsize{DW}}})$ and $k_{\chi,\text{\scriptsize{DW}}}\simeq 0.57 \bar{T}_e^{-1/2}$.
By substituting $\gamma_\text{\scriptsize{DW}}$ and $k_{\chi,\text{\scriptsize{DW}}}$ in Eq.~\eqref{eqn:transport_intermediate}, the cross-field heat flux can be written as
\begin{equation}
    \label{eqn:transport_int}
    q_{\psi,\text{\scriptsize{DW}}}\sim\ 0.2\frac{\bar{T}_e^2\bar{n}}{L_{p,\text{\scriptsize{DW}}}}\,,
\end{equation}
where $\bar{T}_e$ and $\bar{n}$ are evaluated at the LCFS.

We note that $\bar{T}_e$ appearing in Eq.~\eqref{eqn:transport_int} depends implicitly on $L_p$. 
In order to progress, we balance $S_p$ with the parallel losses at the target plates.
In the case of parallel heat transport dominated by convection (the regime of GBS simulations), the global balance in the SOL can be written as
\begin{equation}
    \label{eqn:parallel_balance}
    \int_\text{SOL} \bar{p}_e \bar{c}_s \mathrm{d}l \sim S_p\,,
\end{equation}
where we assume plasma outflowing at the sound speed velocity at the target plates. 
An order of magnitude estimate of $\bar{T}_e$ is then derived by integrating Eq.~\eqref{eqn:parallel_balance}, leading to~\cite{giacomin2021}
\begin{equation}
    \label{eqn:te_lcfs}
    \bar{T}_e \sim \biggl(\frac{5}{4} \frac{S_p}{\bar{n}L_{p,\text{\scriptsize{DW}}}}\biggr)^{2/3}\,.
\end{equation}
By replacing the estimate of $\bar{T}_e$, Eq.~\eqref{eqn:te_lcfs}, into Eq.~\eqref{eqn:transport_int}, the cross-field turbulent heat flux at the LCFS becomes 
\begin{equation}
\label{eqn:transport}
    q_{\psi,\text{\scriptsize{DW}}} \sim\ 0.3 \frac{S_p^{4/3}}{n^{1/3}L_{p,\text{\scriptsize{DW}}}}\,.
\end{equation}

The integral on the right-hand side of Eq.~\eqref{eqn:flux_balance} can be evaluated by assuming $q_{\psi,\text{\scriptsize{DW}}}$ constant along the LCFS, thus leading to 
\begin{equation}
    \label{eqn:balance_approx}
    S_p\sim 2 \pi a\sqrt{\frac{1+\kappa^2}{2}} q_{\psi, \text{\scriptsize{DW}}}\,,
\end{equation}
where $a$ is the tokamak minor radius and $\kappa$ is the plasma elongation at the LCFS.
The analytical estimate of $L_{p,\text{\scriptsize{DW}}}$ is obtained from Eq.~\eqref{eqn:balance_approx} by replacing the analytical estimate of $q_{\psi,\text{\scriptsize{DW}}}$, Eq.~\eqref{eqn:transport}, into Eq.~\eqref{eqn:balance_approx}. This leads to 
\begin{equation}
    \label{eqn:lp_dw}
    L_{p,\text{\scriptsize{DW}}}\sim (1+\kappa^2)^{3/14} a^{3/7} S_p^{1/7} \bar{n}^{-1/7} \,,
\end{equation}
where $\bar{n}$ and $\bar{T}_e$ are evaluated at the LCFS and a numerical factor of order unity is omitted. 

The edge pressure gradient length in Eq.~\eqref{eqn:lp_dw} can also be derived in the limit of parallel heat conduction larger than the parallel heat convection (typical experimental regime) and it is denoted as $L_{p,\text{\scriptsize{DW}}}'$ (the prime symbol is used to distinguish the estimate derived in the heat conduction limit from the heat convection limit).
The global balance in Eq.~\eqref{eqn:parallel_balance} becomes
\begin{equation}
    \label{eqn:parallel_balance_cond}
    S_p\simeq \int_\text{SOL} q_\parallel\, \mathbf{b}\cdot\frac{\nabla \chi}{||\nabla \chi||}\,\mathrm{d}l\,,
\end{equation}
where the parallel heat flux in the SOL is given by
\begin{equation}
    q_\parallel = \chi_{\parallel e} \nabla_\parallel \bar{T}_e = \frac{2}{7}\chi_{\parallel e0} \bar{T}_e^{5/2}\nabla_\parallel \bar{T}_e\,.
\end{equation}
An analytical estimate of the electron temperature at the LCFS can be obtained from Eq.~\eqref{eqn:parallel_balance_cond} by assuming $\nabla_\parallel \sim 1/L_\parallel$, with $L_\parallel$ the parallel connection length in the SOL. This leads to~\cite{stangeby2000}
\begin{equation}
    \label{eqn:twopoint}
    \bar{T}_e \sim \biggl(\frac{7}{2}\frac{S_p L_\parallel}{\chi_{\parallel e0} L_{p,\text{\scriptsize{DW}}}'}\frac{q}{a\rho_*}\biggr)^{2/7}\,,
\end{equation}
where we approximate $\mathbf{b}\cdot \nabla\chi/||\nabla \chi ||\sim q/(\rho_*a)$.

The cross-field turbulent heat flux in the conduction limit is obtained by substituting Eq.~\eqref{eqn:twopoint} into Eq.~\eqref{eqn:transport_int}, which leads to
\begin{equation}
    \label{eqn:transp_cond}
    q_{\psi,\text{\scriptsize{DW}}}' \sim \rho_*^{-4/7} S_p^{4/7} L_\parallel^{4/7} q^{4/7} L_p^{-1} \chi_{\parallel e 0}^{-4/7} L_\parallel^{-4/7} a^{-4/7}\bar{n} \,.
\end{equation}
Finally, by substituting $q_{\psi,\text{\scriptsize{DW}}}'$ in Eq.~\eqref{eqn:balance_approx}, the pressure gradient length in the drift-wave regime and conduction limit is obtained, that is
\begin{equation}
\label{eqn:lp_dw_cond}
    L_{p,\text{\scriptsize{DW}}}'\sim \rho_*^{-4/11}(1+\kappa^2)^{7/22}a^{3/11}S_p^{-3/11}\chi_{\parallel e 0}^{-4/11} L_\parallel^{4/11} q^{4/11} \bar{n}^{7/11}\,.
\end{equation}

\subsection{Resistive ballooning turbulence}

As shown in Sec.~\ref{sec:overview}, both the presence of the electromagnetic fluctuations and the use of the Boussinesq approximation have a weak effect on turbulence and equilibrium profiles in the resistive ballooning regime. 
Therefore, the analysis of this regime, carried out in the electrostatic limit and reported in Ref.~\onlinecite{giacomin2020transp}, remains valid. This analysis leads to an analytical estimate of the equilibrium pressure gradient length near the separatrix in the convective limit, which can be written as
\begin{equation}
    \label{eqn:lp}
    L_{p,\text{\scriptsize{RB}}} \sim \frac{5^{8/17}\pi^{12/17}}{2^{13/17}}\biggl[\rho_*^3\nu_0^6 q^{12} a^{12}(1+\kappa^2)^{6}\bar{n}^{10}S_p^{-4}\biggr]^{1/17}\,.
\end{equation}
We highlight that the theoretical scaling of $L_{p,\text{\scriptsize{RB}}}$ in Eq.~\eqref{eqn:lp} has been successfully validated against a multi-machine database of L-mode discharges, as reported in Ref.~\onlinecite{giacomin2021}.  

In Ref.~\onlinecite{giacomin2022density}, the evaluation of the pressure gradient has been extended to the heat conduction limit. The result is reported here,
\begin{equation}
\label{eqn:lp_cond}
    L_{p,\text{RB}}' \sim \pi^{28/29} \rho_*^{-1/29}(1+\kappa^2)^{14/29} a^{20/29}\nu_0^{14/29}q^{36/29} \bar{n}^{42/29} \chi_{\parallel e 0}^{-8/29} L_\parallel^{8/29} S_p^{-20/29}\,,
\end{equation}
expressed in terms of GBS normalized parameters.

\section{Turbulent transport regime transitions}\label{sec:transitions}

This section is focused on the study of the transitions between the different regimes in the phase space of Fig.~\ref{fig:phase_space}. Three main parameters controlling turbulent transport in the tokamak boundary are identified. 
In addition, theoretical scaling laws that describe the H-mode density limit, the L-mode density limit and the $\beta$ limit of Fig.~\ref{fig:phase_space} are derived in terms of engineering parameters. 

\subsection{H-mode density limit}

As shown in Sec.~\ref{sec:overview}, the drift-wave regime is characterized by a steeper edge pressure gradient than the resistive ballooning regime and, therefore, a higher energy confinement time.
We associate the transition from the drift-wave to the resistive ballooning regime to a H-mode density limit, which typically occurs at high collisionality (the physics behind the L-H transition and the pedestal formation involves kinetic effects,~\cite{stoltzfus2012,dickinson2012,boedo2016} which are not included in the fluid model considered here).

The transition between the drift-wave and the resistive ballooning regimes occurs when ${L_{p,\text{\scriptsize{RB}}} \simeq L_{p,\text{\scriptsize{DW}}}}$, which leads to
\begin{equation}
\label{eqn:hmode_dl_bou}
    \frac{\nu_0}{S_p^{15/14}} \sim \frac{2^{13/6}}{5^{4/3}\pi^2} \rho_*^{-1/2} q^{-2} a^{-11/14} (1+\kappa^2)^{-11/28}\bar{n}^{-29/14}\,.
\end{equation}
We note that the left-hand side of Eq.~\eqref{eqn:hmode_dl_bou} is a function of the parameters $\nu_0$ and $S_p$, which are varied across the simulation set, while the right-hand side is approximately constant in our simulations scan and it is of the order of $10^{-4}$. 
The boundary defined by Eq.~\eqref{eqn:hmode_dl_bou} agrees well with the results of GBS simulations, namely the simulations with $\nu_0/S_p^{15/14}\lesssim 10^{-4}$ are found mainly driven by the drift-wave instability (see Fig.~\ref{fig:projection_phase_space}).

A scaling law of the maximum edge density  that can be achieved in the drift-wave regime is derived from Eq.~\eqref{eqn:hmode_dl_bou}. In physical units, Eq.~\eqref{eqn:hmode_dl_bou} leads to
\begin{equation}
\label{eqn:hdl}
    n_\text{HDL} \sim A^{9/29} P_\text{SOL}^{15/29}a^{-11/29}(1+\kappa^2)^{-11/58}q^{-28/29}R_0^{-22/29}B_T^{11/29}\,,
\end{equation}
where $n_\text{HDL}$ is in units of $10^{20}$~m$^{-3}$, $A$ is the mass number of the main ion species, $a$ is the tokamak minor radius in m, $R_0$ is the tokamak major radius in m, $\kappa$ is the plasma elongation, $q$ is the edge safety factor, $B_T$ is the toroidal magnetic field at the tokamak magnetic axis in T and $P_\text{SOL}$ is the power crossing the separatrix in MW.

Similarly, the scaling law for the H-mode density limit in the heat conduction regime is derived by imposing the condition $L_{p,\text{\scriptsize{RB}}}' \simeq L_{p,\text{\scriptsize{DW}}}'$, with $L_{p,\text{\scriptsize{DW}}}'$ and $L_{p,\text{\scriptsize{RB}}}'$ given by Eqs.~\eqref{eqn:lp_dw_cond}~and~\eqref{eqn:lp_cond}, respectively.
In physical units, this leads to
\begin{equation}
    \label{eqn:hdl_cond}
    n_\text{HDL}' \sim A^{11/37}P_\text{SOL}^{19/37} a^{-19/37} (1+\kappa^2)^{-15/37}    q^{-36/37}R_0^{-22/37}B_T^{15/37}\,,
\end{equation}
where $L_\parallel \sim q R_0$ has been used.
Apart from a stronger dependence on $\kappa$ in Eq.~\eqref{eqn:hdl_cond}, the two scaling laws share a similar dependence on the engineering parameters.

Although the Greenwald density and the H-mode density limit, Eqs.~\eqref{eqn:greenwald}~and~\eqref{eqn:hdl}, are associated with different transitions, it is useful to compare them. 
By making the plasma current dependence explicit, Eq.~\eqref{eqn:hdl} is written as
\begin{equation}
\label{eqn:hdl_Ip}
    n_\text{HDL}' \sim A^{11/37} P_\text{SOL}^{19/37}R_0^{14/37}B_T^{-21/37}(1+\kappa^2)^{-15/37}\frac{I_p^{36/37}}{a^{91/37}}\,.
\end{equation}
The H-mode density limit scaling in Eq.~\eqref{eqn:hdl_Ip} shares with the Greenwald scaling the main dependence on $I_p$ and $a$, but also depends on $P_\text{SOL}$ and $B_T$. 
A recent empirical scaling of the H-mode density limit obtained from a log-linear regression applied to ASDEX Upgrade H-mode density limit data shows a relatively strong dependence on the heating power, i.e. $n_\text{HDL} \propto P_\text{heat}^{0.4}$,\cite{bernert2014h} which agrees well with the power dependence shown in Eq.~\eqref{eqn:hdl_Ip}. On the other hand, the H-mode density limit scaling in Eq.~\eqref{eqn:hdl_Ip} shows a stronger dependence on $I_p$ and $B_T$ than the one reported in Ref.~\onlinecite{bernert2014h}, where $n_\text{HDL} \propto B_T^{-0.3}I_p^{0.6}$.
We note that the power dependence of the H-mode density limit is still subject of discussion. For example, Refs.~\onlinecite{mertens2000},~\onlinecite{borrass2004}~and~\onlinecite{huber2013} report no or weak power dependence in the H-mode density limit.

\subsection{L-mode density limit}

The results of GBS simulations presented in Sec.~\ref{sec:overview} show that electromagnetic perturbations at high collisionality have no effect on turbulence and equilibrium profiles at the tokamak boundary, if $\beta_{e0}$ is below the $\beta$ limit. 
Therefore, the results derived in Ref.~\onlinecite{giacomin2020transp} in the electrostatic limit are valid also when electromagnetic effects are considered.
Following Refs.~\onlinecite{giacomin2020transp,giacomin2022density}, the crossing of the density limit can be associated with a collapse of the edge pressure gradient due to  enhanced turbulent transport. 
This collapse is estimated by assuming that $L_p$ becomes comparable to a significant fraction of the tokamak minor radius, i.e.  $L_p \sim a$.
By imposing this condition in Eq.~\eqref{eqn:lp}, we obtain
\begin{equation}
\label{eqn:res_lim_fin}
    \frac{\nu_0^{3/2}}{S_p}\sim \frac{2^{13/4}}{25\pi^3}\frac{a^{5/4}}{\rho_*^{3/4}q^3\bar{n}^{5/2}(1+\kappa^2)^{3/2}}\,.
\end{equation}
The left-hand side of Eq.~\eqref{eqn:res_lim_fin} depends on the parameters $\nu_0$ and $S_p$, which are varied across the simulation set, while the right-hand side is approximately constant in all the simulations considered here and is approximately equal to 0.5.  
As shown in Fig.~\ref{fig:projection_phase_space}, the theoretical limit provided by Eq.~\eqref{eqn:res_lim_fin} agrees well with the results of GBS electromagnetic simulations.
In fact, turbulent eddies in the simulations with $\nu_0^{3/2}/S_p \gtrsim 0.5$ have a radial extension comparable to the tokamak minor radius, $1/(k_\psi a) \simeq 0.5$, and lead to a very large cross-field turbulent transport and, consequently, to a flat pressure profile.  

Similarly to the H-mode density limit, Eq.~\eqref{eqn:res_lim_fin} is written in physical units and in terms of engineering parameters, leading to
\begin{equation}
\label{eqn:den_lim}
    n_{\text{DL}} \sim A^{-1/10} a^{1/2} B_T^{6/5}P_\text{SOL}^{2/5} q^{-6/5} R_0^{-7/10} (1+\kappa^2)^{-3/5}\,,
\end{equation}
where $n_{\text{DL}}$ is the maximum achievable edge density in units of $10^{20}$~m$^{-3}$, $a$ and $R_0$ are the tokamak minor and major radii in m, $B_T$ is the toroidal magnetic field in T, $P_\text{SOL}$ is the power crossing the separatrix in units of MW, $k$ is the plasma elongation at the LCFS.
On the other hand, in the limit of parallel heat conduction larger than parallel heat convection, the density limit scaling can be written as
\begin{equation}
    \label{eqn:den_lim_cond}
    n_\text{DL}' \sim A^{1/6} a^{3/14} P_\text{SOL}^{10/21} R_0^{-43/42} q^{-22/21}(1+\kappa^2)^{-1/3}B_T^{2/3}\,.
\end{equation}
The density limit scaling in Eq.~\eqref{eqn:den_lim_cond} has been validated against a multi-machine database in Ref.~\onlinecite{giacomin2022density}.

In order to compare Eq.~\eqref{eqn:den_lim_cond} to the empirical scaling in Eq.~\eqref{eqn:greenwald}, we rewrite Eq.~\eqref{eqn:den_lim_cond} in terms of the plasma current, 
\begin{equation}
    \label{eqn:den_lim_ip}
     n_\text{DL}' \sim  A^{1/6} P_\text{SOL}^{10/21}R_0^{1/42}B_T^{-8/21}(1+\kappa^2)^{-1/3}\frac{I_p^{22/21}}{a^{79/42}}\,.
\end{equation}
We note that Eqs.~\eqref{eqn:greenwald}~and~\eqref{eqn:den_lim_ip} share a main dependence on $I_p$ and $a$, but the density limit in Eq.~\eqref{eqn:den_lim_ip} depends on $P_\text{SOL}$, in agreement with experimental observations.~\cite{stabler1992,mertens1997,rapp1999,esposito2008,huber2013}

\begin{figure}
    \centering
    \includegraphics[scale=0.8]{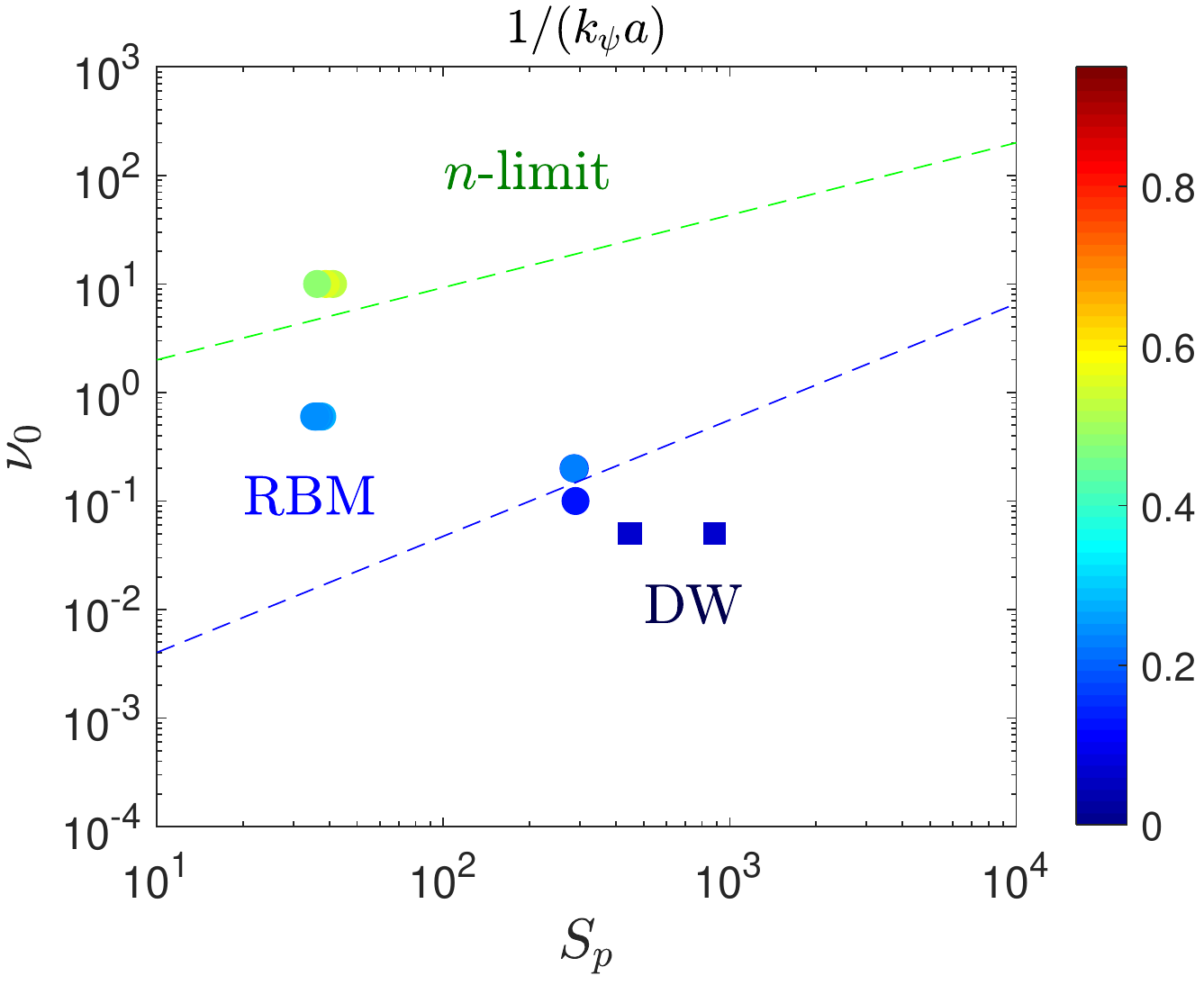}
    \caption{H-mode (blue line) and L-mode (green line) density limit, Eqs.~\eqref{eqn:hmode_dl_bou}~and~\eqref{eqn:res_lim_fin} respectively, represented on the phase space identified by the parameters $\nu_0$ and $S_p$. The region enclosed between the two transitions represent the regime for a stable operation in L-mode. }
    \label{fig:hl_dl}
\end{figure}

We compare now the H-mode and L-mode density limits. Fig.~\ref{fig:hl_dl} shows the analytical estimates of these two boundaries (see  Eqs.~\eqref{eqn:hmode_dl_bou}~and~\eqref{eqn:res_lim_fin}) on the phase space defined by the parameters $\nu_0$ and $S_p$. 
The region in Fig.~\ref{fig:hl_dl} between these two boundaries corresponds to a stable L-mode operation beyond the H-mode density limit. Although this region appears quite wide in terms of GBS parameters, its area can be significantly smaller in experiments. 
In fact, the comparison between the theoretical scaling in Eq.~\eqref{eqn:den_lim_cond} and a multi-machine database, reported in Ref.~\onlinecite{giacomin2022density}, shows the presence of a numerical factor in Eq.~\eqref{eqn:den_lim_cond}, which reduces the region of stable L-mode plasma between the L-mode and the H-mode density limits in Fig.~\ref{fig:hl_dl}.

\subsection{$\beta$-limit}

The regime transition observed at high $\beta$ (red plane in Fig.~\ref{fig:phase_space}), which we denote as $\beta$-limit, is associated with the onset of the ideal ballooning instability that becomes the dominant instability when the parameter $\alpha_{\text{MHD}}$, defined in Eq.~\eqref{eqn:alpha_mhd_first}, exceeds a value of the order of unity.~\cite{lortz1978,Zeiler1997}
We note that $\alpha_\text{MHD}$ depends on $L_p$ and $\bar{T}_e$, which in turn depend on turbulent transport. 
Since the $\beta$-limit is approached from the resistive ballooning regime by increasing $\beta_{e0}$, we consider $L_p$ as the result of the resistive ballooning transport, given by Eq.~\eqref{eqn:lp}.
The electron temperature at the LCFS is then estimated by using Eq.~\eqref{eqn:te_lcfs}. 
By substituting the analytical estimates of $\bar{T}_e$  and $L_{p,\text{\scriptsize{RB}}}$ into Eq.~\eqref{eqn:alpha_mhd_first}, the criterion for the onset of an ideal mode is written as 
\begin{equation}
\label{eqn:alpha_mhd}
    \alpha_\text{MHD} \sim \frac{\beta_{e0}}{2^{1/17}5^{2/17}\pi^{20/17}} \biggl(\frac{q^{14}S_p^{18}}{\rho_*^{22}a^{20}(1+\kappa^2)^{10}\nu_0^{10}\bar{n}^{11}}\biggr)^{1/17} \gtrsim 1\,,
\end{equation}
which leads to
\begin{equation}
\label{eqn:beta_limit}
    \frac{\beta_{e0}S_p^{18/17}}{\nu_0^{10/17}}\gtrsim 2^{1/17}5^{2/17}\pi^{20/17}\biggl(\frac{\rho_*^{22}a^{20}(1+\kappa^2)^{10}\bar{n}^{11}}{q^{14}}\biggr)^{1/17}\,,
\end{equation}
where the left-hand side in Eq.~\eqref{eqn:beta_limit} depends on the parameters $\beta_{e0}$, $\nu_0$ and $S_p$, which are varied across the simulation scan, while the right-hand side is approximately equal to 0.2 for all the simulations.
As shown in Fig.~\ref{fig:phase_space}, the radial extension of the turbulent eddies in simulations with $\beta_{e0}S_p^{18/17}/\nu_0^{10/17}\gtrsim 0.2$ is approximately equal to the tokamak minor radius. The whole plasma confined region is therefore characterized by the presence of large scale and large amplitude fluctuations, leading to a total loss of plasma and heat  (see Fig.~\ref{fig:density_2d}~(d)).

A scaling law for the appearance of the ideal modes in engineering parameters is obtained by writing Eq.~\eqref{eqn:alpha_mhd} in physical units. This leads to
\begin{equation}
    \label{eqn:alpha_phys}
    \alpha_\text{MHD} \sim 0.1\; A^{1/17} P_\text{SOL}^{18/17}q^{14/17}a^{-20/17}B_T^{-14/17}n^{-11/17}R^{-6/17}(1+\kappa^2)^{-10/17}\gtrsim 1\,,
\end{equation}
where $a$ and $R_0$ are the tokamak minor and major radii in m, $B_T$ is the toroidal magnetic field in T, $P_\text{SOL}$ is the power crossing the separatrix in units of MW, $\kappa$ is the plasma elongation at the LCFS and $n$ is the edge density in units of $10^{20}$~m$^{-3}$.

Similarly to the H-mode and L-mode density limit, the $\beta$-limit is provided also in the heat conduction limit,
\begin{equation}
    \label{eqn:alpha_phys_cond}
    \alpha_\text{MHD}' \sim 0.2\; A^{9/29}P_\text{SOL}^{34/29} q^{20/29} a^{-36/29}B_T^{-22/29}n^{-25/29}R_0^{-16/29}(1+\kappa^2)^{17/29}\gtrsim 1\,.
\end{equation}
The major difference between Eq.~\eqref{eqn:alpha_phys} and Eq.~\eqref{eqn:alpha_phys_cond} stems from the $\kappa$ dependence, beside the stronger dependence on $A$ and $R_0$ in Eq.~\eqref{eqn:alpha_phys_cond}.

\section{Remarks on the edge turbulence phase space and comparison with past investigations}\label{sec:comparison}

We analyse here the main analogies and differences between the edge phase space outlined in Fig~\ref{fig:phase_space} and the one derived in Ref.~\onlinecite{rogers1998} that, based on the results of flux-tube simulations, has constituted the paradigm to explain the edge turbulent regimes for more than two decades. 
We also compare our phase space of edge turbulence to the one recently derived in Ref.~\onlinecite{eich2021} in terms of the electron density and electron temperature at the separatrix.

The first important difference between the phase space in Fig.~\ref{fig:phase_space} and the one in Ref.~\onlinecite{rogers1998} stems from the edge parameters that delineate the phase space.
The parameters chosen in Ref.~\onlinecite{rogers1998}, $\alpha_\text{MHD}$ and $\alpha_d$, depend on $L_p$, which in turn depends on turbulent transport. A constant value of $L_p$ is considered in Ref.~\onlinecite{rogers1998} across the different regimes. 
However, the simulations presented in this work clearly show a dependence of $L_p$ on collisionality, heat source and $\beta$. 
This dependence is retained in our phase space of edge turbulence through the analytical estimates of $L_p$ derived for both the drift-wave and resistive ballooning driving instabilities (see Eqs.~\eqref{eqn:lp_dw}~and~\eqref{eqn:lp}).
We note that in Ref.~\onlinecite{eich2021} the parameter $\alpha_d$ is replaced by $\alpha_t = (L_p/R_0)^{1/2}/(\pi \alpha_d)^2\propto \nu$, which retains the key dependence on the plasma collisionality and removes the dependence on $L_p$.

In agreement with the phase space of Ref.~\onlinecite{rogers1998}, Fig.~\ref{fig:projection_phase_space} shows the presence of a regime of reduced turbulent transport at low collisionality, i.e. high value of $\alpha_d$, where the drift-wave instability dominates over the resistive ballooning instability. In Ref.~\onlinecite{rogers1998}, this regime of reduced transport is associated with the H-mode of tokamak operation, while here it is associated with a regime near the H-mode density limit at high collisionality, where a fluid model can be applied. 
Therefore, the transition from the drift-wave regime to the resistive ballooning regime is claimed to correspond to the H-mode density limit.
On the other hand, the phase space in Ref.~\onlinecite{eich2021} identifies the transition to a drift-wave dominated regime, where flow shear suppresses turbulence, with the L-H transition.

The density limit presented here significantly differs from the one derived in Ref.~\onlinecite{rogers1998}.
In fact, in the phase space of Ref.~\onlinecite{rogers1998}, the density limit can be achieved only for values of $\alpha_\text{MHD}$ larger than 0.1 and it is fundamentally linked to electromagnetic effects. 
The importance of electromagnetic effects in the density limit is also highlighted in Ref.~\onlinecite{eich2021}, which associates the crossing of the density limit with a transition from the electrostatic to the electromagnetic resistive ballooning regime, a transition that leads to a strong increase of turbulent eddy size and, therefore, to an extremely large turbulent transport.
However, the simulations presented here show that the density limit can be crossed at any value of $\beta_{e0}$, and even in the electrostatic limit, with turbulent transport that becomes extremely large also in the absence of electromagnetic modes. 
In fact, the size of turbulent eddies in the proximity of the density limit crossing can be very large independently of the presence of electromagnetic modes.
This can be seen by balancing the interchange drive term, $2 C(p_e)$, and the parallel current term, $\nabla_\parallel j_\parallel$, in Eq.~\eqref{eqn:vorticity}. 
The term $C(p_e)$ is estimated from the linearized pressure equation, which is obtained by linearizing and summing Eqs.~\eqref{eqn:density}~and~\eqref{eqn:electron_temperature}, i.e.
\begin{equation}
    \gamma \tilde{p}_e \sim i \rho_*^{-1}\frac{\bar{p}_e}{L_p} k_\chi \tilde{\phi}\,, 
\end{equation}
where $\gamma\simeq \sqrt{2\bar{T}_e/(\rho_* L_p)}$ is the growth rate of the interchange instability~\cite{mosetto2013} and $k_\chi$ denotes the corresponding poloidal wave vector. 
This leads to
\begin{equation}
\label{eqn:curv_pressure}
    C(\tilde{p}_e) \sim \frac{\bar{p}}{\gamma\rho_*L_p}k_\chi^2\tilde{\phi}\,.
\end{equation}
The term $\nabla_\parallel j_\parallel$ is estimated from the electron parallel momentum balance, Eq.~\eqref{eqn:electron_velocity}, that, linearized, leads to
\begin{equation}
\label{eqn:linear_ele_parallel}
\gamma\psi  \sim \nu \tilde{j}_\parallel + \nabla_\parallel \tilde{\phi}\,,    
\end{equation}
where the electron inertia is neglected.
The term on the left-hand side of Eq.~\eqref{eqn:linear_ele_parallel} is estimated by using Eq.~\eqref{eqn:ampere}, leading to $\gamma \psi \sim \gamma \beta_{e0} /(2 k_\perp^2) \tilde{j}_\parallel$. 
For typical values of electron density and electron temperature at the separatrix of a discharge in the proximity of the density limit, $n_e\simeq 5\times 10^{19}$~m$^{-3}$ and $T_e\simeq 30$~eV,  the ratio of $\nu$ to $\gamma \beta_{e0} /(2 k_\perp^2)$ is of the order of 10. 
Consequently, in Eq.~\eqref{eqn:linear_ele_parallel} the term $\nu j_\parallel$  dominates over the term $\partial\psi/\partial t$, and the resistive and the parallel electric field terms balance. 
As a consequence, taking its parallel divergence, Eq.~\eqref{eqn:linear_ele_parallel} can be written as
\begin{equation}
\label{eqn:nabla_par_j}
\nabla_\parallel \tilde{j}_\parallel \sim \frac{\nabla_\parallel^2 \tilde{\phi}}{\nu}\,.  
\end{equation}
Equations~\eqref{eqn:curv_pressure}~and~\eqref{eqn:nabla_par_j} lead to $k_\chi \propto \nu^{-1/2}$. Namely, the size of turbulent eddies increases with resistivity, becoming very large even in absence of electromagnetic modes, and can be ascribed to a change of the linear properties of the driving resistive ballooning modes.
As an aside, we note that the term $\partial \psi/\partial t$ may dominate over the term $\nu j_\parallel$ at low collisionality and high $\beta$ in the drift-wave regime.

Dedicated experimental investigations have been carried out in the past with the aim of validating the phase space derived in Ref.~\onlinecite{rogers1998} (see, e.g., Refs.~\onlinecite{eich2020,labombard2005}). In particular, experimental observations show that turbulent transport in the tokamak boundary strongly depends on $\alpha_d$, especially at high density, pointing out the important role played by the edge collisionality in the density limit,~\cite{eich2020,labombard2001,labombard2005} in agreement with the phase space derived here.
On the other hand, the boundary of the density limit experimentally found in Ref.~\onlinecite{labombard2005} shows also a dependence on the $\alpha_\text{MHD}$ parameter, a result that may suggest a role played by electromagnetic fluctuations.
However, we remark that $\alpha_\text{MHD}$ depends on the edge pressure gradient and, therefore, on turbulent transport, independently of its electrostatic or electromagnetic nature, i.e. a relation between the density limit and the $\alpha_\text{MHD}$ parameter is not sufficient to conclude that the density limit is caused by electromagnetic rather than electrostatic turbulent transport. 
In addition, the pressure gradient dependence appearing in both $\alpha_\text{MHD}$ and $\alpha_d$ makes these two parameters correlated. Therefore, they cannot be varied independently experimentally, thus making it challenging to decouple the effects due to collisionality and the ones due to $\beta$.

\section{Conclusions}\label{sec:conclusions}

The results of three dimensional, flux-driven, global, electromagnetic turbulent simulations, carried out by using the GBS code avoiding the Boussinesq approximation,  are used to identify the phase space of plasma turbulence and transport in the tokamak boundary.
Based on the results of these simulations, four turbulent transport regimes are identified: (i) a regime at intermediate values of collisionality, heat source and $\beta$, where turbulence is driven by resistive ballooning modes, which is associated with the standard L-mode of tokamak operation; (ii) a regime at low collisionality, large heat source and intermediate values of $\beta$, where turbulence is mainly driven by the drift-wave instability,  associated with the H-mode tokamak operation at high density; (iii) a regime of extremely large turbulent transport, where turbulence is driven by resistive ballooning modes, which is associated with the crossing of the L-mode density limit; and (iv) a regime at large values of $\beta$, associated with the crossing of the $\beta$ limit, where the ideal ballooning instability drives turbulence, generating large scale modes that affect the entire confined region and lead to a total loss of plasma and heat. 
In addition, the transition from the drift-wave to the resistive ballooning regime is associated with the H-mode density limit.

The electromagnetic simulations considered here point out a weak effect of electromagnetic fluctuations on turbulence and equilibrium profiles for realistic $\beta$ values that are below the $\beta$ limit. 
In particular, the results presented here show that the density limit can be achieved independently of the value of $\beta$, thus with a secondary role played by electromagnetic fluctuations.
In addition, the comparison of the GBS simulations presented here to the ones reported in Ref.~\onlinecite{giacomin2020transp} shows that the Boussinesq approximation has a strong effect on turbulence and equilibrium profiles at low collisionality, while no significant effect related to the use of the Boussinesq approximation is observed at intermediate and high collisionality. 

Analytical scaling laws of the H-mode and L-mode density limit as well as of the $\beta$ limit are derived and compared to the results of GBS simulations, showing an overall good agreement. 
These scaling laws are also provided in terms of engineering parameters, thus allowing for a direct application to the experiments.
We highlight that both the H-mode and L-mode density limit scaling laws depend on the power crossing the separatrix, which will be significantly larger in future fusion devices than in present day tokamaks. 
The scaling law of the L-mode density limit in Eq.~\eqref{eqn:den_lim_cond} has been recently validated against a multi-machine database in Ref.~\onlinecite{giacomin2022density}, predicting a factor two higher density limit for ITER than the corresponding prediction based on the Greenwald density limit scaling.
On the other hand, a prediction of the ITER H-mode density limit based on Eq.~\eqref{eqn:hdl_cond} requires first a detailed validation with current experiments. Therefore, the results of the present work call for a comparison between the H-mode density limit scaling in Eq.~\eqref{eqn:hdl_cond} against a multi-machine database of H-mode density limit discharges.

\section*{Acknowledgments}

The authors thank T. Eich, A. Pau and O. Sauter for useful discussions.
Discussions within the framework of the TSVV1 EUROfusion project led by T. Goerler are also gratefully acknowledged.
The simulations presented herein were carried out in part at the Swiss National Supercomputing Center (CSCS) under the project IDs s882 and s1028, in part on the CINECA Marconi supercomputer under the GBSedge and LHPED21 projects and in part using the JFRS-1 supercomputer system at Computational Simulation Centre of International Fusion Energy Research Centre (IFERC-CSC) in Rokkasho Fusion Institute of QST (Aomori, Japan).
This work, supported in part by the Swiss National Science Foundation, was carried out within the framework of the EUROfusion Consortium and has received funding from the Euratom research and training programme 2014 - 2018 and 2019 - 2020 under grant agreement No 633053. The views and opinions expressed herein do not necessarily reflect those of the European Commission. 

\appendix 
\renewcommand{\thesection}{}
\section{Shear flow effects}
\renewcommand{\thesection}{\Alph{section}}
\renewcommand{\thefigure}{A.\arabic{figure}}
\setcounter{figure}{0}

In order to assess the impact of the $\mathbf{E}\times\mathbf{B}$ mean sheared flow on the linear properties of the ballooning and drift-wave instabilities and, consequently, on the H-mode density limit, we consider a reduced physical model derived from Eqs.~\eqref{eqn:density}--\eqref{eqn:ampere}, 
\begin{align}
\label{eqn:linear_first}
    \frac{\partial}{\partial t} \nabla\cdot (n \nabla_\perp \phi) &= -\rho_*^{-1} \nabla \cdot \bigl[\phi, n\nabla_\perp \phi\bigr]+2C(p_e)+\nabla_\parallel \nabla_\perp^2\psi\,,\\
    \frac{\partial}{\partial t}\Bigl(\frac{\beta_{e0}}{2}-\frac{m_e}{m_i n}\nabla_\perp^2\Bigr)\psi &=\nu \nabla_\perp^2\psi + \nabla_\parallel\phi - \frac{1.71}{n} \nabla_\parallel p_e - 1.71\frac{\beta_{e0}\rho_*^{-1}}{2}\bigl[\psi, n\bigr]\,,\\
\label{eqn:linear_last}
    \frac{\partial }{\partial t} p_e &= -\rho_*^{-1}\bigl[\phi,p_e] + \nabla_\parallel \nabla_\perp^2\psi\,,
\end{align}
which avoids the use of the Boussinesq approximation and accounts for electromagnetic effects and $\mathbf{E}\times\mathbf{B}$ sheared flows.
The physical model in Eqs.~\eqref{eqn:linear_first}--\eqref{eqn:linear_last} is  linearized by assuming $\nabla_\parallel \sim 1/q$ and $\phi(r,\theta) = \phi_0(r) + \phi_1(r)\exp(i m \theta)$, with $\phi_1/\phi_0 \ll 1$, and similarly for all other fields.
Eqs.~\eqref{eqn:linear_first}--\eqref{eqn:linear_last} are solved numerically by considering an equilibrium $\phi_0 = \tanh[(r-r_0)/L_\phi] - 1$, $\psi_0 = 0$, $p_{e0} = 1-\tanh[(r-r_0)/L_p]$, $n_0 = 1-\tanh[(r-r_0)/L_n]$,  $m_i/m_e = 2000$, $\beta_{e0}=10^{-4}$, $\rho_*=0.002$ and $r_0=150$. 
In particular, the growth rate $\gamma$ and the poloidal wave number $m$ are computed for different values of $\nu_0$, $L_\phi$ and $L_n$, with $L_p=L_n/2$.
An implicit equation for $L_p$ is obtained by imposing a balance between perpendicular and parallel transport, i.e. $q_\psi/L_p \sim q_\parallel/L_\parallel$, where $q_\psi$ is given by Eq.~\eqref{eqn:transport_intermediate} and $L_\parallel \sim q\rho_*^{-1}$. This leads to~\cite{Ricci2013} 
\begin{equation}
    \label{eqn:lp_num}
    L_p\sim q \biggl(\frac{\gamma}{k_\chi}\biggr)_\text{max}\,.
\end{equation}

\begin{figure}[t]
    \centering
    \subfloat[Full model]{\includegraphics[width=0.32\textwidth]{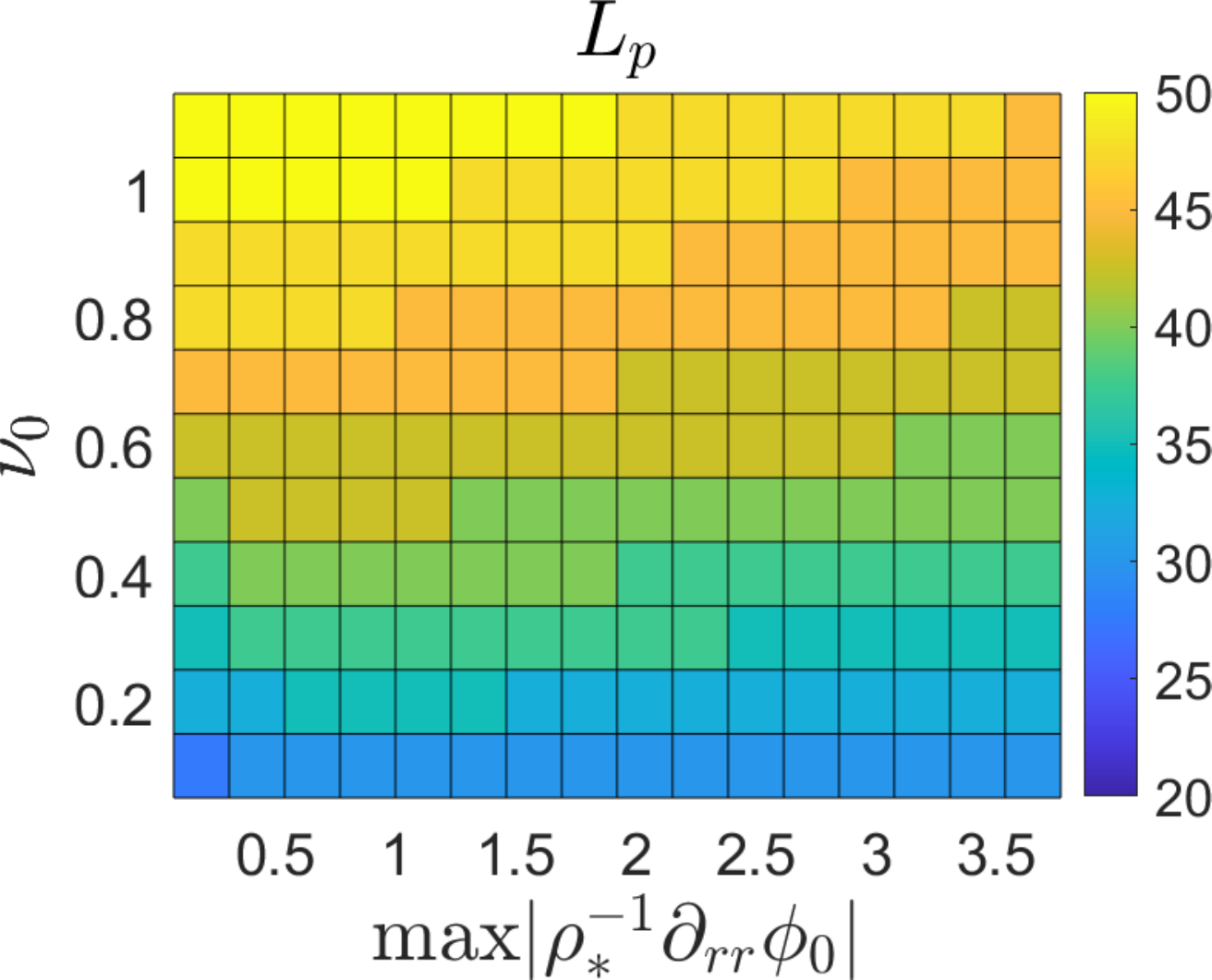}}\,
    \subfloat[No drift-wave]{\includegraphics[width=0.32\textwidth]{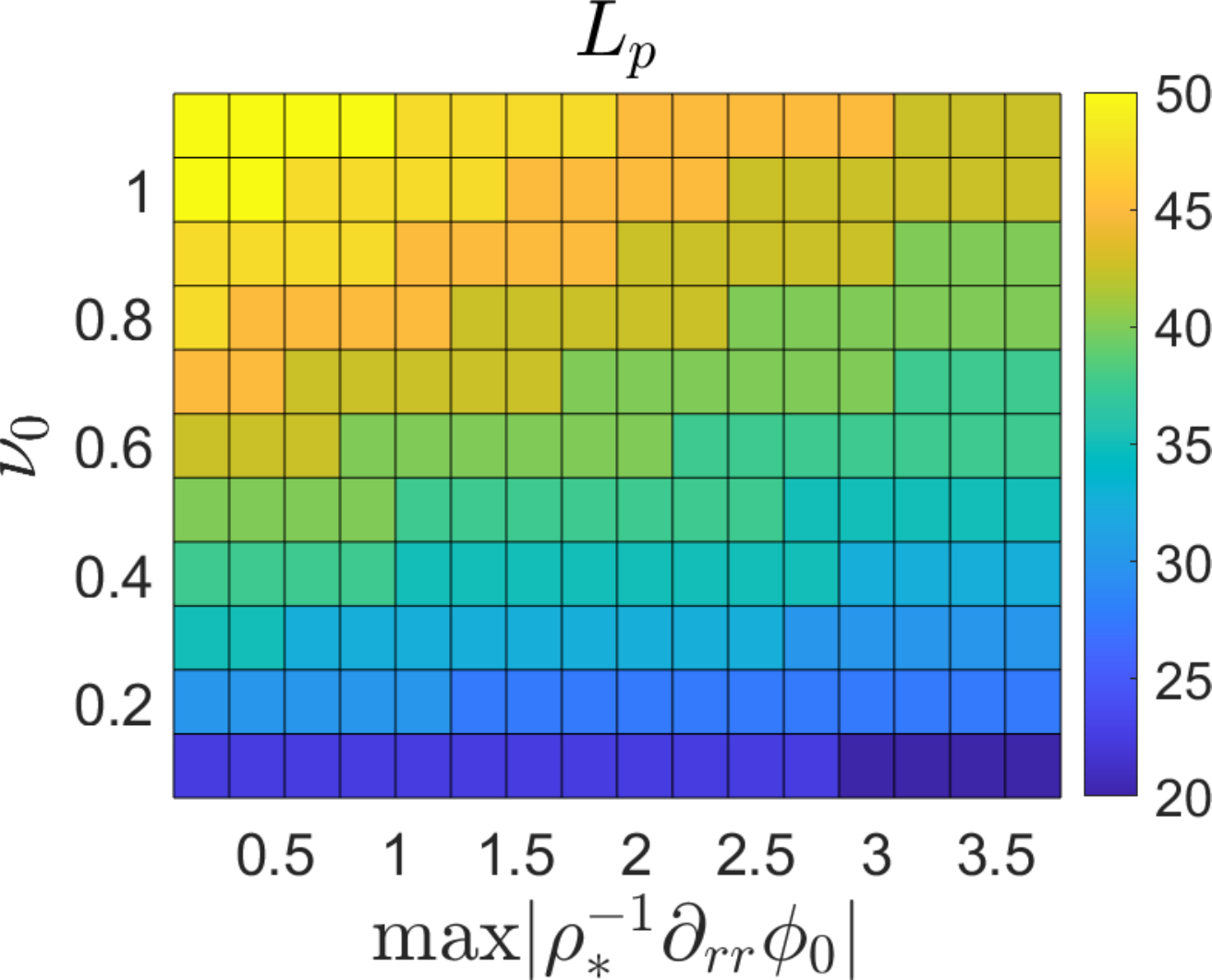}}\,
    \subfloat[No ballooning]{\includegraphics[width=0.32\textwidth]{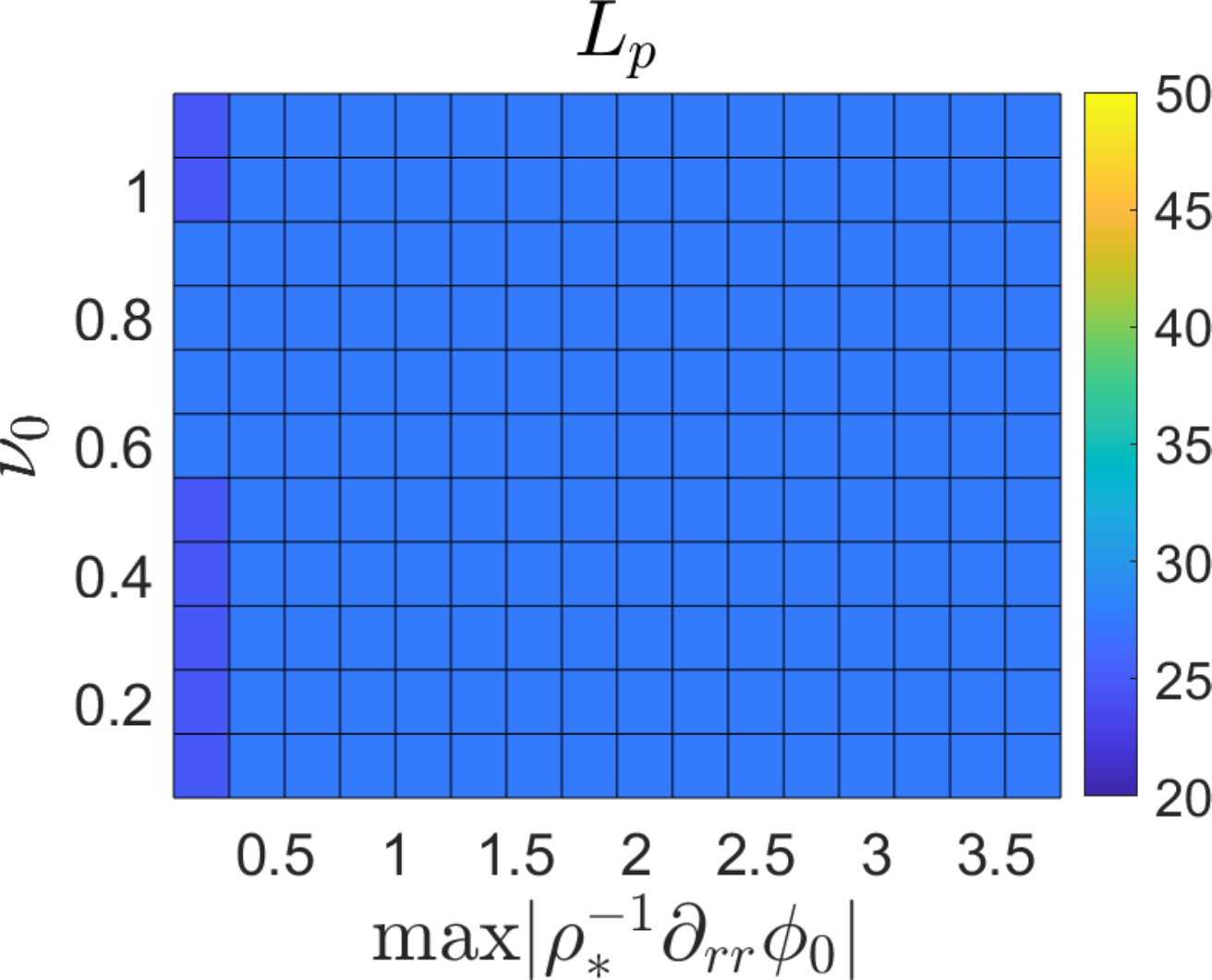}}
    \caption{Numerical solution of Eq.~\eqref{eqn:lp_num} at different values of resistivity, $\nu_0$, and $\mathbf{E}\times\mathbf{B}$ shear rate, $\mathrm{max}|\rho_*^{-1}\partial_{rr}\phi_0|$, when considering the full model in Eqs.~\eqref{eqn:linear_first}--\eqref{eqn:linear_last} (a), the model without the drift-wave instability (b) or the model without the ballooning instability (c). }
    \label{fig:lp_linear}
\end{figure}

The numerical solution of Eq.~\eqref{eqn:lp_num}  at different values of resistivity, $\nu_0$, and $\mathbf{E}\times\mathbf{B}$ shear rate, evaluated as $\mathrm{max}|\rho_*^{-1}\partial_{rr}\phi_0|$, is shown in Fig.~\ref{fig:lp_linear}. 
At high values of $\nu_0$, the resistive ballooning instability dominates and, consequently, $L_p$ decreases with $\nu_0$, in agreement with Eq.~\eqref{eqn:lp}. 
In addition, $\gamma/k_\chi$ decreases as the $\mathbf{E}\times\mathbf{B}$ shear rate increases, and this reduces the value of $L_p$. 
On the other hand, at low values of $\nu_0$, the numerical solution of Eq.~\eqref{eqn:lp_num} becomes independent of $\nu_0$ and $\mathrm{max}|\rho_*^{-1}\partial_{rr}\phi_0|$, and reaches a minimum. This corresponds to a transition to a regime where turbulence is driven by the drift-wave instability.
This is shown by removing the drift-wave instability from Eqs.~\eqref{eqn:linear_first}--\eqref{eqn:linear_last}.
In this case $L_p$ decreases with $\nu_0$ also at small values of $\nu_0$, as shown in Fig.~\ref{fig:lp_linear}~(b), reaching values that are smaller than the ones obtained from the solution of the full system. 
On the other hand, a very weak dependence on $\nu_0$ is observed when the ballooning instability is removed (see Fig.~\ref{fig:lp_linear}~(c)), in agreement with the analytical estimate of $L_{p,\text{\scriptsize{DW}}}$ in Eq.~\eqref{eqn:lp_dw}, which is independent of $\nu_0$. 
We also note that $L_p$ in Fig.~\ref{fig:lp_linear}~(c)  depends very weakly on the shear rate. 
Therefore, the effect of the $\mathbf{E}\times\mathbf{B}$ mean sheared flow can be neglected when parameters in the proximity of the H-mode density limit are considered, thus justifying the use of Eq.~\eqref{eqn:lp_dw}, which is derived under the assumption of negligible mean sheared flows.

\bibliographystyle{unsrt}
\bibliography{library}

\end{document}